\documentclass[traditabstract]{aa} % for the abstract without structuration 
\usepackage{graphicx}

\usepackage{txfonts}

\usepackage{natbib}

\usepackage{amstext}
\usepackage[verbose]{placeins}
\usepackage{tikz}
\usepackage{array}
\usepackage{amssymb}
\usepackage{url} 
\newcommand{\sersic}{S\'{e}rsic}	
\newcommand{\bd}{\begin{displaymath}}
\newcommand{\ed}{\end{displaymath}}
\newcommand{\be}{\begin{equation}}
\newcommand{\ee}{\end{equation}}
\newcommand{\beaa}{\begin{eqnarray*}}
\newcommand{\eeaa}{\end{eqnarray*}}
\newcommand{\bea}{\begin{eqnarray}}
\newcommand{\eea}{\end{eqnarray}}

\def\ourlens{RX\,J1131$-$1231{}}

\def\zd{z_{\rm d}}
\def\zs{z_{\rm s}}

\begin{document}
   \title{GLaD: Gravitational Lensing and Dynamics, combined analysis to unveil properties of high-redshift galaxies}

  \titlerunning{GLaD: Gravitational Lensing and Dynamics}

   \author{G. Chiriv\`i\inst{1,2}
   		 \and    
          A. Y$\rm \i$ld$\rm \i$r$\rm \i$m\inst{1}
                    \and
          S. H. Suyu\inst{1,2,3}  
                  \and 
          A. Halkola        
          }

   \institute{Max-Planck-Institut f\"ur Astrophysik, Karl-Schwarzschild Str. 1, 85741 Garching, Germany\\
              \email{chirivig@MPA-Garching.MPG.DE}
         \and 
         Physik-Department, Technische Universit\"at M\"unchen, James-Franck-Stra\ss{}e~1, 85748 Garching, Germany
           \and 
          Academia Sinica Institute of Astronomy and Astrophysics (ASIAA), 11F of ASMAB, No.1, Section 4, Roosevelt Road, Taipei 10617, Taiwan 
           }

   \date{Received --; accepted --}

% \abstract{}{}{}{}{} 
% 5 {} token are mandatory
 
  \abstract
  % context heading (optional)
  % {} leave it empty if necessary  
  {Dynamical modelling of Integral-Field-Unit (IFU) stellar kinematics is a powerful tool to unveil the dynamical structure and mass build-up of galaxies in the local Universe, while gravitational lensing is nature's cosmic telescope to explore the properties of galaxies beyond the local Universe. We present a new approach which unifies dynamical modelling of galaxies with the magnification power of strong gravitational lensing, to reconstruct the structural and dynamical properties of high-redshift galaxies. By means of axisymmetric Jeans modelling, we create a dynamical model of the source galaxy, assuming a surface brightness and surface mass density profile. We then predict how the source’s surface brightness and kinematics would look like when lensed by the foreground mass distribution and compare with the mock observed arcs of strong gravitational lensing systems. For demonstration purposes, we create and analyse mock data of the strong lensing system \ourlens. By modelling both the lens and source, we recover the dynamical mass within the effective radius of strongly lensed high-redshift sources within 5\% uncertainty, and we improve the constraints on the lens mass parameters by up to 50\%. This machinery is particularly well suited for future observations from large segmented-mirror telescopes, such as the James Webb Space Telescope, that will yield high sensitivity and angular-resolution IFU data for studying distant and faint galaxies. }

% We use axisymmetric Jeans modelling to create a dynamical model of the source galaxy, assuming a surface brightness and surface mass density profile, and then predict how the source’s surface brightness and kinematics would look like when lensed into arcs. 
%SHS: By reconstructing the distorted kinematic data and comparison with the observed arcs of real strong gravitational lensing systems, we are capable of constraining the lens and source properties with high precision. 
%By reconstructing the distorted kinematic data and comparing with the observed arcs of real strong gravitational lensing systems, we can constrain the lens and source properties with high precision. 
%
%SHS: This machinery is aiming for data from future telescopes such as JWST, as the sensitivity and resolution of state-of-the-art telescopes is insufficient. 
%This machinery is particularly well suited for future observations from telescopes such as the James Webb Space Telescope that will yield high sensitivity and angular-resolution IFU data for studying distant and faint galaxies.
%
%For demonstration purposes, we mock up the strongly lensed source galaxy of the strong lensing system \ourlens\, and map the stellar kinematics onto the lens plane. By modelling both the spatially-resolved stellar kinematics and the foreground lens mass distribution, 
%SHS: we're able to recover
%we recover 
%
%the structural and dynamical properties of strongly lensed high-redshift sources within 
%SHS: up to 5\% uncertainty.}
%5\% uncertainty.}
%
  % aims heading (mandatory)
   {}
  % methods heading (mandatory)
   {}
  % results heading (mandatory)
   {}
  % conclusions heading (optional), leave it empty if necessary 
   {}


   \maketitle
%
%________________________________________________________________

\section{Introduction}
\label{sec:intro}

The progenitors of today's massive galaxy population are thought to be small and dense \citep{Daddi2005,Trujillo2006,Zirm2007,vanderWel2008,vanderWel2014,vanDokkum2008,Szomoru2010,SzomoruFranxvanDokkum2012}, disky \citep{Toft2005,Trujillo2006,vanderWel2011,Chang2013}, with quenched star formation and old stellar populations \citep{Kriek2006,Kriek2008,Kriek2009,Toft2007,Cimatti2008,vanDokkum2010}, small \sersic\ indices and high stellar velocity dispersion \citep{vanDokkumKriekFranx2009,Bezanson2011,Toft2012,vandeSande2013} \citep[see][for a review]{Cappellari2016}. How they came to be nowadays' most massive galaxies, enclosing most of the stellar mass in the Universe \citep{Fukugita1998,Hogg2002,Bell2003,Baldry2004}, is still a debated topic. Significant progress on the numerical side now suggests a two-phase scenario for the formation and evolution of the massive galaxy population \citep{2010ApJ...725.2312O,2016MNRAS.456.1030W}, while our comprehension on the observational side has developed immensely by means of Integral-Field Unit (IFU) observation.  These have proven to be a groundbreaking tool to unveil the structural and dynamical properties of galaxies \citep[see][for a review]{Cappellari2016}. In fact, Integral Field Spectroscopy (IFS) has become an essential tool in astrophysics, allowing us to obtain a spectrum in every spaxel on the sky covering the entire galaxy field, and therefore obtaining, as outcoming data, a 3D data cube \citep[][]{Bacon1995,deZeeuw2002}. %Cappellari2011a,Bundy2015,Cappellari2016,Cappellari&Copin2003}. 
From these data cubes we can reconstruct the stellar and gas kinematic maps, which are important tracers of the underlying gravitational potential due to visible (stars and gas) and non-visible (dark) matter, allowing us to explore the matter content, matter distribution and internal dynamics of galaxies, which hold clues regarding their assembly history. A way in which these information can be quantified is via dynamical modelling, that allows a detailed description of the galaxy's dynamics. Key results were obtained using dynamical models fitted to stellar and gas kinematics, as for example mass determinations of supermassive black holes in galaxies \citep[e.g.][]{Gebhardt2000,Barth2001,Cappellari&Verolme2002,Sarzi2001,Gultekin2009a,Gultekin2009b,Gultekin2012,VanDenBosch2010,McConnell2012,VanDenBosch2012,Walsh2013,Thomas2016}, the determination of their stellar mass-to-light ratios, dark matter fractions, total mass profiles and slopes \citep[e.g.][]{Kronawitter2000,Weijmans2009,Murphy2011,Cappellari2013,Cappellari2015,Yildirim2017}. The revelation of fast and slow rotators \citep{Emsellem2007} and how these are potentially linked to the formation and evolution histories of today’s most massive galaxies are also important findings \citep{Naab2014}. However, obtaining resolved gas and stellar kinematic maps is mainly possible for galaxies in the local Universe.\\
To unveil the properties of galaxies in the high-redshift Universe, we can exploit natural cosmic telescopes such as gravitational lenses. Gravitational lensing is a relativistic effect for which the light travelling from a source towards the observer is bent by the presence of matter (baryonic and dark) in between. Consequently, the source will be observed at a different position than it actually is, distorted in shape, and in some cases also multiply imaged (in the so-called $``$strong lensing$"$ regime). It will also appear magnified by a factor $\mu$ (magnification). By modelling the lens mass distribution, one can reconstruct the image positions and magnification $\mu$ of the source and, thanks to surface brightness conservation, 
%SHS: the intrinsic brightness of the source galaxy can be reconstructed. 
%SHS note: omit intrinsic since *intrinsic* SB cannot be reconstructed uniquely due to mass sheet degeneracy
the surface brightness distribution of the source galaxy can be reconstructed.
This will allow one to study high-redshift galaxies, providing crucial probes of structure formation and galaxy evolution \citep[e.g.][among others]{Oldham2017}. Combined lensing and stellar dynamic techniques were employed in previous works \citep[e.g.][among others]{Treu&Koopmans2002,Treu&Koopmans2004,BarnabeKoopmans2007,Koopmans2009,vandeVen2010,Barnabe2011,Barnabe2012}, mainly to study properties of lens galaxies.\\
In this paper we present GLaD (Gravitational Lensing and Dynamics), a software that is able to unify dynamical modelling of galaxies with the magnification power of strong gravitational lensing, to reconstruct the dynamical properties of high-redshift source galaxies. GLaD is able to model the source stellar kinematics using axisymmetric Jeans modelling, assuming a source surface brightness and surface mass density profile, and then predict how these maps will look like when lensed and distorted by a strong gravitational lens. This allows us to compare the predicted arcs and images directly to the observed strong gravitational lensing systems, in order to provide improved constraints for both the source and the deflector. \\
Previous works have applied similar methods exploiting the gravitational lensing magnification to study spatially resolved kinematics of background sources \citep[e.g.][]{Jones2010,Dye2015,Rybak2015,Swinbank2015,Newman2017,Newman2018,DiTeodoro2018,Girard2018,Patricio2018,Rizzo2018}. %SHS: revised previous phrase to use plural form
 However, these methods mostly rely on a pre-modelled lens model, which is kept fixed during the dynamical analysis. This technique is suboptimal for it does not
%SHS: allow to quantify 
quantify
degeneracies between the lens mass and source kinematics. Another trait which is common to these works is performing the dynamical analysis on the source plane, by $``$de-lensing$"$ the kinematic map from the lens plane to the source plane. A disadvantage of this approach is the induced pixel correlation derived from the $``$de-lensing$"$. Moreover, it is not clear how to properly characterise noise properties on the source plane. Finally, the resolution on the source plane is dependent on differential magnification, which is an effect that must be taken into account. Some 
%SHS: of these works 
recent works
overcame these issues partially or totally \citep{Patricio2018,Rizzo2018}, but differ from our method for the use of different dynamical modelling techniques and scientific goal, since they focus on the star-forming population by tracing the gas kinematics, which are not necessarily a pure tracer of the gravitational potential. Moreover, and unlike previous methods, we do not rely on a pixellated source reconstruction but instead assume parametrised profiles for the source, from which we can easily recover source properties of interest such as its total mass, ellipticity, \sersic\ index and effective size. \\
The paper is organised as follows: we describe our method for both the lensing and the dynamical analysis in Section \ref{sec:glad_method}. We present a test case based on mock data of \ourlens\ and show how well we are able to recover the lensing and dynamical parameters using GLaD in Section \ref{sec:rxj1131}. We discuss and conclude in Section \ref{sec:conclude}. Throughout the paper, parameter constraints are given by the median
values with the uncertainties given by the 16th and 84th percentiles (corresponding to 68\% credible intervals (CI)) of the
marginalised probability density distributions.  We assume a flat
$\Lambda$CDM cosmology with $H_0=70 \rm{\, km\, s^{-1}\, Mpc^{-1}}$ and
$\Omega_{\Lambda}=1-\Omega_{\rm M}=0.73$.  From the redshifts of the lens and the
source galaxies in Section \ref{sec:rxj1131}, one arcsecond at the lens (source) 
plane in \ourlens\ corresponds to $4.43\ (7.04) {\rm \, kpc}$.

%__________________________________________________________________

\section{GLaD methodology}
\label{sec:glad_method}
GLaD is a software developed to combine the stellar dynamical and strong lensing analyses with the aim of reconstructing the light and mass properties of the source galaxy. We use axisymmetric Jeans modelling to create a dynamical model of the source galaxy assuming a mass and a surface brightness distribution, and then predict how the source's surface brightness and kinematics would look like when lensed into arcs. We then compare our predictions of the source's distorted surface brightness and projected second order velocity moment $\overline{\varv_{\mathrm{LOS}}^{2}}$ with that of the observed arcs. In this method, we simultaneously fit the source and deflector properties, consisting of the source and deflector's light and mass distributions, the source orbital anisotropy parameter $\beta$ and inclination $i$. We use Bayesian analysis to infer the best-fit parameter values together with their uncertainties and degeneracies. In this Section we explore the different functionalities of GLaD: we introduce the lens profiles we employ in our analysis (Section \ref{subsec:MassLightProfiles}), we describe the dynamics and lensing analysis (Section \ref{subsec:jam_mod} and \ref{subsec:glee}) and explain how these are combined to obtain a consistent joint analysis (Section \ref{subsec:comb_likel}). 
\subsection{Mass and light profiles}
\label{subsec:MassLightProfiles}
To describe the source and lens galaxy's mass distribution, we use simply parametrised profiles. We describe the source's total mass distribution with a singular pseudoisothermal elliptical mass
distribution \citep[PIEMD;][]{KassiolaKovner93} with dimensionless
surface mass density
\begin{equation}
\label{eq:kappaPIEMD}
\kappa_{\rm piemd}(x,y)\Big|_{\rm z_s=\infty}= \frac{\theta_{\rm E,\infty}}{2 \sqrt{R_{\rm em}^2 +r_{\rm c}^2}},  
\end{equation}
where $(x,y)$ are the coordinates in the galaxy's plane (along the semi-major and semi-minor axes), $ R_{\rm em} $ is the elliptical mass radius,
\begin{equation}
\label{eq:Rem}
R_{\rm em}= \sqrt[]{\frac{x^2}{(1+e)^2} + \frac{y^2}{(1-e)^2}}, 
\end{equation}
$e$ is the ellipticity $e= \frac{1-q}{1+q}$ with $q$ the axis ratio,  $\theta_{\rm E, \infty}$ is the lens Einstein radius for source at redshift infinity and $r_{\rm c}$ is the core radius.  The mass distribution is then suitably rotated by its position angle $\theta$ and shifted by the centroid position of the coordinate system used. The parameters that identify this profile are its centroid position $(x_{\rm c}, y_{\rm c})$, its axis ratio $q$, its position angle $\theta$, its Einstein radius $\theta_{\rm E, \infty}$, and its core radius $r_{\rm c}$.\\
To represent the mass distribution of the lens we use a softened power-law elliptical mass distribution \citep[SPEMD; ][]{Barkana98} which contains an additional parameter as compared to the PIEMD profile, which is the slope $\gamma$. Its convergence, is 
\begin{equation}
\label{eq:kappaSPEMD}
\kappa_{\rm spemd}(x,y)\Big|_{\rm z_s=\infty}= \theta_{\rm E,\infty} \left( x^2+\frac{y^2}{q^2}+r_{\rm c}^2 \right) ^{-\gamma},  
\end{equation}
where $q$ is the axis ratio, $r_{\rm c}$ is the core radius, $\gamma$ is the power law index, which is 0.5 for an isothermal profile  \citep{Barkana98}. 
To the lens mass we also add a constant external shear, described by the lens potential parametrised by
\be
\label{eq:extsh}
\psi_{\rm ext}(\vartheta,\varphi)=\frac{1}{2} \gamma_{\rm ext, \infty}
\vartheta^2 \cos(2(\varphi-\phi_{\rm ext})),
\ee
where $\vartheta$ and $\varphi$ are polar coordinates such that
$x=\vartheta \cos(\varphi)$ and $y=\vartheta
\sin(\varphi)$,  $\gamma_{\rm ext, \infty}$ is the shear strength for source at redshift infinity and $\phi_{\rm ext}$ is the shear angle ($\phi_{\rm ext}=0\degr$ corresponds to a shearing along the $x$-direction while $\phi_{\rm ext}=90\degr$ corresponds to a shearing along the $y$-direction). The shear centre is arbitrary since it is not observable. Finally, to disentangle the baryonic mass from the dark component, we alternatively model the lens mass with a composite mass model. The composite model consists of multiple aforementioned PIEMD profiles (see ``chameleon'' profile below), to account for the contribution of the stellar density, and a \citet*[NFW]{Navarro1997} profile for the dark matter density distribution
\begin{equation}
\rho (r) =  \frac{\rho_\text{s}}{\frac{r}{r_\text{s}} \left( 1 + \frac{r}{r_\text{s}} \right)^{2}},
\label{eq:nfw}
\end{equation}
where $\rho_\text{s}$ is the characteristic overdensity and $r_\text{s}$ is the scale radius. \\
To represent both the lens and source light profiles we use a \sersic\ profile, whose intensity is given by 
\begin{equation}
\label{eq:Sersic}
 I(R_{\rm q})= I_{\rm e} \exp \Bigg\{ -k\  \bigg[ \bigg(\frac{R_{\rm q}}{R_{\rm eff}}\bigg)^{\frac{1}{n}}-1 \bigg] \Bigg\},  
\end{equation}
where $k$ is approximately $2n-\frac{1}{3}$, $R_{\rm q}=\sqrt{x^2+y^2/q^2}$ and $n$ is the \sersic\ index which, for most galaxies, spans values between $\frac{1}{2}< n < 10$ and whose value is correlated to the size and the magnitude of the galaxy. The \sersic\ profile is defined by the centroid position $x_{\rm c,s}$ and $y_{\rm c,s}$, the axis ratio $q$, the position angle $\theta$, the \sersic\ amplitude $I_{\rm e}$, the effective radius $R_{\rm eff}$ and the \sersic\ index $n$. In addition to the \sersic\ profile, we also employ the so-called ``chameleon'' profile, which is a profile that mimics the \sersic\ and allows analytic computations of lensing quantities \citep[e.g.,][]{Maller2000, Dutton2011, Suyu2014}. This profile is composed of a difference of two isothermal profiles, namely
\bea
\label{eq:isothermal}
\displaystyle	
{\Large L(x,y)}  = & {\Large \frac{L_0}{1+q_\text{l}}}  {\Large \left( \frac{1}{\sqrt{x^2 + y^2 / q_\text{l}^2 + 4 w_\text{c}^2 / (1+q_\text{l})^2}} \right.} \nonumber  \\
& { \Large - \left. \frac{1}{\sqrt{x^2 + y^2 / q_\text{l}^2 + 4 w_\text{t}^2 / (1+q_\text{l})^2}} \right)~.}
\eea
where $q_\text{l}$ is the axis ratio, $w_\text{t}$ and $w_\text{c}$ are parameters of the profile (with $w_\text{t}>w_\text{c}$ to keep $L>0$). To represent the baryonic mass we scale the chameleon light profile by a constant mass-to-light ratio.\\

\subsection{Dynamical Modelling: Axisymmetric Jeans Modelling}
\label{subsec:jam_mod}

\begin{figure*}[h!]
  \centering
  \includegraphics[trim=10 10 10 10, clip,width=1.\textwidth]{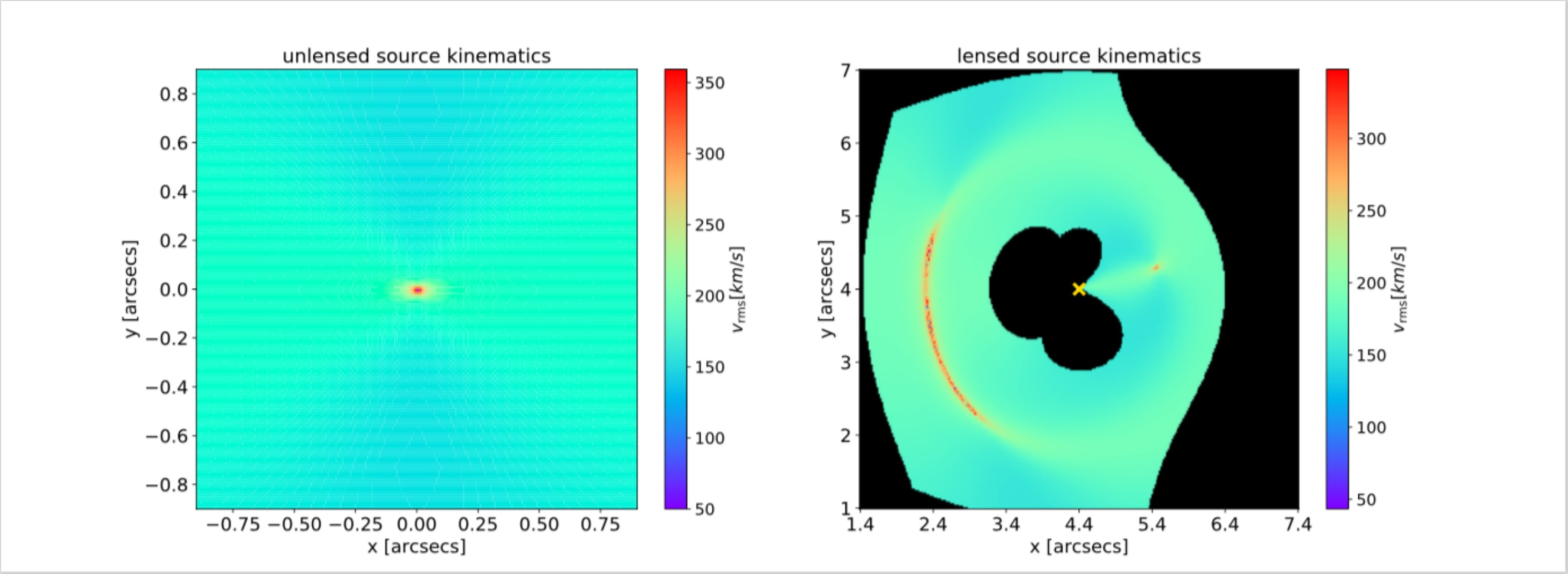}
  \caption{Simulated source stellar kinematics with GLaD. We produce an axisymmetric Jeans model of a mock source galaxy on the source plane (left panel) and we predict the lensed $\varv_{\rm rms}$ profile in the image plane (right panel). The masked black region corresponds to regions in the image plane which have no correspondence on the source plane. The yellow cross marks the lens centroid position.}
  \label{fig:kin_ex}
\end{figure*}

The distribution function (DF) $f(\vec{x},\vec{\varv})$ describing the positions $\vec{x}= (x,y,z)$ and velocities $\vec{\varv}= (\varv_{x},\varv_{y},\varv_{z})$ of a large system of stars must satisfy the fundamental equation of stellar dynamics, the steady-state collisionless Boltzmann equation (CBE) \citep{Binney&Tremaine1988}. With the axial symmetry assumption (in 3 dimensions), multiplication of the CBE with powers of the velocity moment, and subsequent integration over velocity space, we can reduce the Boltzmann equation into the Jeans equations \citep{Jeans1922}, written in terms of the cylindrical coordinates $(R,z, \phi)$
\be
\label{eq:jeans1}
\scalebox{1.3}{$ \frac{\nu \overline{\varv^{2}_{R}}-\nu \overline{\varv^{2}_{\phi}}}{R}  +  \frac{\partial \left( \nu \overline{\varv^{2}_{R}} \right)}{\partial R}  + \frac{ \partial \left( \nu \overline{\varv_{R}} \overline{\varv_{z}} \right)}{\partial z} = -\nu \frac{\partial \Phi}{\partial R}$,}
\ee
\be
\label{eq:jeans2}
\scalebox{1.3}{$ \frac{\nu \overline{\varv_{R}} \overline{\varv_{z}}}{R}  +  \frac{\partial \left( \nu \overline{\varv^{2}_{z}} \right)}{\partial z}  + \frac{ \partial \left(\nu \overline{v_{R}} \overline{\varv_{z}} \right) }{\partial R} = -\nu \frac{\partial \Phi}{\partial z}$,}
\ee
where $\Phi$ is the gravitational potential, $\nu$ the luminosity density and
\be
\label{eq:jeans_notation}
\scalebox{1.0}{$ \nu \overline{\varv_{i}} \overline{\varv_{j}} \equiv \int{ \varv_{i} \varv_{j} f {\rm d}^{3} \vec{\varv}}$.}
\ee
The axisymmetry assumption seems to be valid, to first order, for most elliptical galaxies, unless photometric or kinematic evidence for bars or triaxiality is present. However, equations (\ref{eq:jeans1}) and (\ref{eq:jeans2}) are still quite general and do not uniquely specify a solution \citep{Cappellari2008}. By specifying the shape and orientation of the intersection of the velocity ellipsoid everywhere in the meridional plane, one can uniquely solve the Jeans equations numerically and recover the motion of stars in a gravitational potential $\Phi$ \citep{Cappellari2008}. To solve the Jeans equations under the above assumptions, we use the Jeans Anisotropic MGE \footnote{Jampy and MGEfit, online available via \url{https://www-astro.physics.ox.ac.uk/~mxc/software/}} routine \citep{Cappellari2002,Cappellari&Copin2003, Cappellari2008}. This software makes the assumptions that: (1) the velocity ellipsoid is aligned with the cylindrical coordinate system $(R,z, \phi)$ and (2) the anisotropy is constant and quantified by $\overline{\varv_{R}^{2}}= \beta\overline{\varv^{2}_{z}} $. Using these assumptions together with the boundary conditions that $\nu \overline{\varv}_{z} =0$ for $z \rightarrow \infty$, one can finally solve the Jeans equations for $\overline{\varv_{z}^{2}}$ and $\overline{\varv_{\phi}^{2}}$.\\
To derive solutions for the Jeans equations (\ref{eq:jeans1}) and (\ref{eq:jeans2}), we first parametrise the source galaxy's stellar surface brightness and surface mass density with a Multi-Gaussian Expansion (MGE). The MGE method was initially conceived by \citet{Bendinell1991} and then further developed in other works \citep{MonnetBacon&Emsellem1992,EmsellemMonnet&Bacon1994A,EmsellemDejonghe&Bacon1999,Cappellari2002}. It consists of a series of expansions of the galaxy's image using 2D Gaussians functions
\be
\label{eq:mge}
\scalebox{1.0}{$ I(\tilde{R},\tilde{\theta})=\displaystyle\sum_{i=1}^{N} \frac{L_{i}}{2\pi \sigma^{2}_{i} q_{i}} \exp{\left[ - \frac{1}{2\sigma^{2}_{i}} \left( x^{2} + \frac{y^{2}}{q^{2}_{i}} \right) \right] },$}
\ee
where $N$ is the number of Gaussians with luminosity $L_{i}$, axis ratio $q_{i}$, and dispersion $\sigma_{i}$, $(x,y)$ are a system of coordinates on the plane of the sky centered on the galaxy's nucleus, and $(\tilde{R},\tilde{\theta})$ the relative polar coordinates $(x=\tilde{R}\sin({\tilde{\theta}-\psi_{i}}), y=\tilde{R}\cos({\tilde{\theta}-\psi_{i}}))$ with $\psi_{i}$ the position angle measured counterclockwise from the $y$-axis to the major axis of the Gaussian. All Gaussians are assumed to have the same centre and position angle. This method allows for a straightforward and analytically convenient expression of an arbitrary surface brightness and surface mass density distribution \citep{Cappellari2002}. Since galaxies have an unknown inclination $i$, one needs to deproject the MGE surface brightness and surface mass density profile to get the intrinsic tracer and mass density $\nu$ and $\Phi$ in equations (\ref{eq:jeans1}) and (\ref{eq:jeans2}). Despite not being able to eliminate the intrinsic degeneracy of the deprojection, the MGE method provides realistic densities, resembling real galaxies, when projected at any angle \citep{Cappellari2008}.  With the deprojected MGE of the surface brightness and surface mass density profile, one can readily solve the Jeans equations and perform the integral along the line-of-sight to obtain the total observed first and second order velocity moment
\be
\label{eq:vrms}
\scalebox{1.0}{$\varv_{\rm rms}=\sqrt{\varv^{2}+\sigma_{\rm v}^2},$}
\ee
where $\varv$ and $\sigma_{\rm v}$ are the source galaxy's projected velocity and velocity dispersion respectively. As it is not clear how the ordered and random motions contribute to a particular $\varv_{\rm rms}$ profile a priori, we stick to the prediction of projected second order moment in equation (\ref{eq:vrms}) for our modelling purposes and show an example of the predicted velocity moments ($\varv_{\rm rms}$) simulated with GLaD in Figure \ref{fig:kin_ex}.

\subsection{Gravitational Lensing: GLEE}
\label{subsec:glee}
\begin{figure*}[h!]
  \centering
  \includegraphics[trim=50 0 50 00, clip,width=\textwidth]{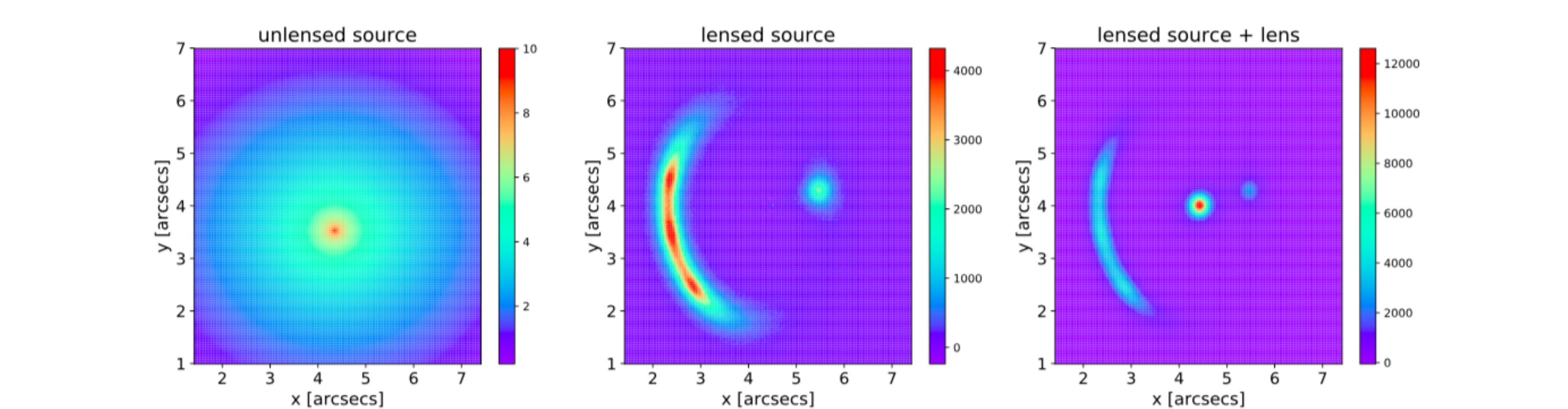}
  \caption{Simulated strong lensing data with GLaD. We model the light of the source (left panel) and the mass and light of the lens and we predict how the surface brightness of the lensed source (central panel) and lensed source with lens (right panel) will look like. The units of the surface brightness are in counts.}
  \label{fig:lensing_ex}
\end{figure*}

Our lensing analysis is carried out using GLEE, a software developed by A. Halkola and S. H. Suyu  \citep{SuyuHalkola2010, Suyu2012a}. This software uses parametrised mass profiles (discussed in Section \ref{subsec:MassLightProfiles}) to describe the different lensing components, such as dark matter halos and galaxies, and allows us to compute the deflection angle for these profiles to map between the source and the image planes. The deflection angle is calculated as
\be
\label{eq:alpha}
\hat{\alpha}(\xi) = \frac{ 4G}{c^{2}} \int_{}^{} \rm d^{2} \xi' \Sigma(\vec{\xi'}) \frac{\vec{\xi}-\vec{\xi'}}{{|\vec{\xi}-\vec{\xi'}|}^2}
\ee
where $ G$ is the gravitational constant, $\xi$ is the 2-dimensional impact vector, and $\Sigma$ is the surface mass density of the galaxy, i.e. the mass density projected onto a plane perpendicular to the incoming light ray, defined as $\Sigma=\kappa\ \Sigma_{\rm crit}$, where $\kappa$ is the dimensionless surface mass density (convergence) and the critical surface mass density is $\Sigma_{\rm crit}= \frac{c^{2}}{4\pi G} \frac{D_{\rm s}}{D_{\rm d}D_{\rm ds}}$, which is a function of the angular diameter distances of lens and source. \\
For the development of GLaD we use only these features of GLEE, namely we use GLEE to (1) compute the deflection angles of analytic mass distributions, (2) model surface brightness profiles of galaxies, and (3) set up multi-lens-plane modelling.  GLaD further builds upon this by incorporating dynamical modelling of the sources. Therefore, a key difference between GLEE and GLaD is on the model of the source: GLEE reconstructs the source surface brightness on a grid of pixels \citep{Suyu2006}, whereas GLaD uses analytic profiles for both the source surface brightness and mass distributions. An example of the use of GLEE to construct the source and lens surface brightnesses is shown in Figure \ref{fig:lensing_ex}.
 
\subsection{Joint Analysis}
\label{subsec:comb_likel}

To sample the parameter space efficiently and obtain the posterior probability distributions we use {\sc Emcee} \citep{ForemanMackey16}, a stable, well tested Python implementation of the affine-invariant ensemble sampler for Markov Chain Monte Carlo (MCMC) proposed by \cite{Goodma&Weare2010}. The posterior probability of the model parameters, collectively denoted by $\vec{\eta}$,  is obtained using Bayes' Theorem 

\begin{equation}
\label{bayesthm}
$$ P(\vec{\eta} | \vec{X}_{\rm obs}) \propto \overbrace{\mathcal{L}(\vec{X}_{\rm obs}|\vec{\eta})}^{\rm likelihood} \overbrace{P(\vec{\eta})}^{\rm prior},$$
\end{equation}
where $P(\vec{\eta})$ is the prior probability on the model parameters, that we always assume to be uniform, and $\vec{X}_{\rm obs}$ collectively denotes our observables (surface brightness $\vec{I}_{\rm obs}$ and velocity map $\vec{\varv}^{\rm obs}_{\rm rms}$). To perform the joint analysis, we combine the lensing likelihood $\mathcal{L}_{\rm lens}$ with the dynamics likelihood $\mathcal{L}_{\rm dyn}$ such that the total likelihood function would be 
\be
\label{eq:like_tot}
\mathcal{L}_{\rm tot}= \mathcal{L}_{\rm lens} \times \mathcal{L}_{\rm dyn}.
\ee
The lensing likelihood is 
\be
\label{eq:like_lens}
\mathcal{L}_{\rm lens}(\vec{I}_{\rm obs} |\vec{\eta}) \propto \exp
  {\left[-\frac{1}{2} \displaystyle\sum_{i=1}^{N_{\rm pixels}}
      \frac{\vert{I}_{i}^{\rm
          obs}-{I}_{i}^{\rm
          pred}(\vec{\eta})\vert^2}{\sigma_{i}^2} \right]},
\ee
where $N_{\rm pixels}$ is the number of pixels in the image, ${I}_{i}^{\rm obs}$ is the observed surface brightness in a certain pixel $i$, 
$I_{i}^{\rm pred}(\vec{\eta})$ is the modelled surface brightness of that same pixel and $\sigma_{i}$ is the uncertainty on that pixel. Here, $I$ and $\sigma$ have the units of counts. We assume our uncertainty in every pixel $\sigma_{i}$ to be composed of the background noise (with $\sigma_{\rm back}$) and by the Poisson noise, namely
\be
\label{eq:sigma}
\sigma_{i}= \sqrt{\sigma_{\rm back}^{2} + I^{\rm obs}_{i}}.
\ee
The dynamics likelihood is
\be
\label{eq:like_dyn}
\mathcal{L}_{\rm dyn}(\vec{\varv}^{\rm obs}_{\rm rms} |\vec{\eta}) \propto \exp
  {\left[-\frac{1}{2}\displaystyle\sum_{i=1}^{N_{\rm bin}}
      \frac{\vert{\varv}_{\rm rms,\it i}^{\rm
          obs}-{\varv}_{\rm rms,\it i}^{\rm
          pred}(\vec{\eta})\vert^2}{\sigma_{{\rm rms},i}^2} \right]}.
\ee
where ${\varv}_{\rm rms,\it i}^{\rm obs}$ is the observed value of Equation (\ref{eq:vrms}) in each bin, ${\varv}_{\rm rms,\it i}^{\rm pred}$ our model prediction, and $\sigma_{{\rm rms},i}$ the error on each bin. To obtain the value of the latter, we assume that the error in each IFU pixel scales inversely proportional to its S/N \citep[see][]{Emsellem&Cappellari2004}. Since we do not mock up the spectroscopic data, we rely on the surface brightness information of the imaging data for assessing the S/N in each IFU spaxel. When binning the $\varv_{\rm rms}$, the signal to noise in each bin is
\be
\label{eq:SN_dyn}
\frac{S}{N}=   \frac{\sum_{j=1}^{\rm N_{\rm pixels}} I_{j}} {\sqrt{\sum_{j=1}^{\rm N_{\rm pixels}} \left( \sigma_{\rm back,\it j}^{2} + I_{j} \right)}}.
\ee
where $\rm N_{\rm pixels}$ is the number of pixels in each bin and $I$ and $\sigma$ here have the units of counts. The error on the ${\varv}_{\rm rms}$ value in each bin is
\be
\label{eq:err_dyn}
\sigma_{{\rm rms}}= \frac{{\varv}_{\rm rms}}{\frac{S}{N}}.
\ee

\section{Demonstration: \ourlens}
\label{sec:rxj1131}

\begin{figure}
  \includegraphics[trim=0 0 0 0, clip,width=\columnwidth]{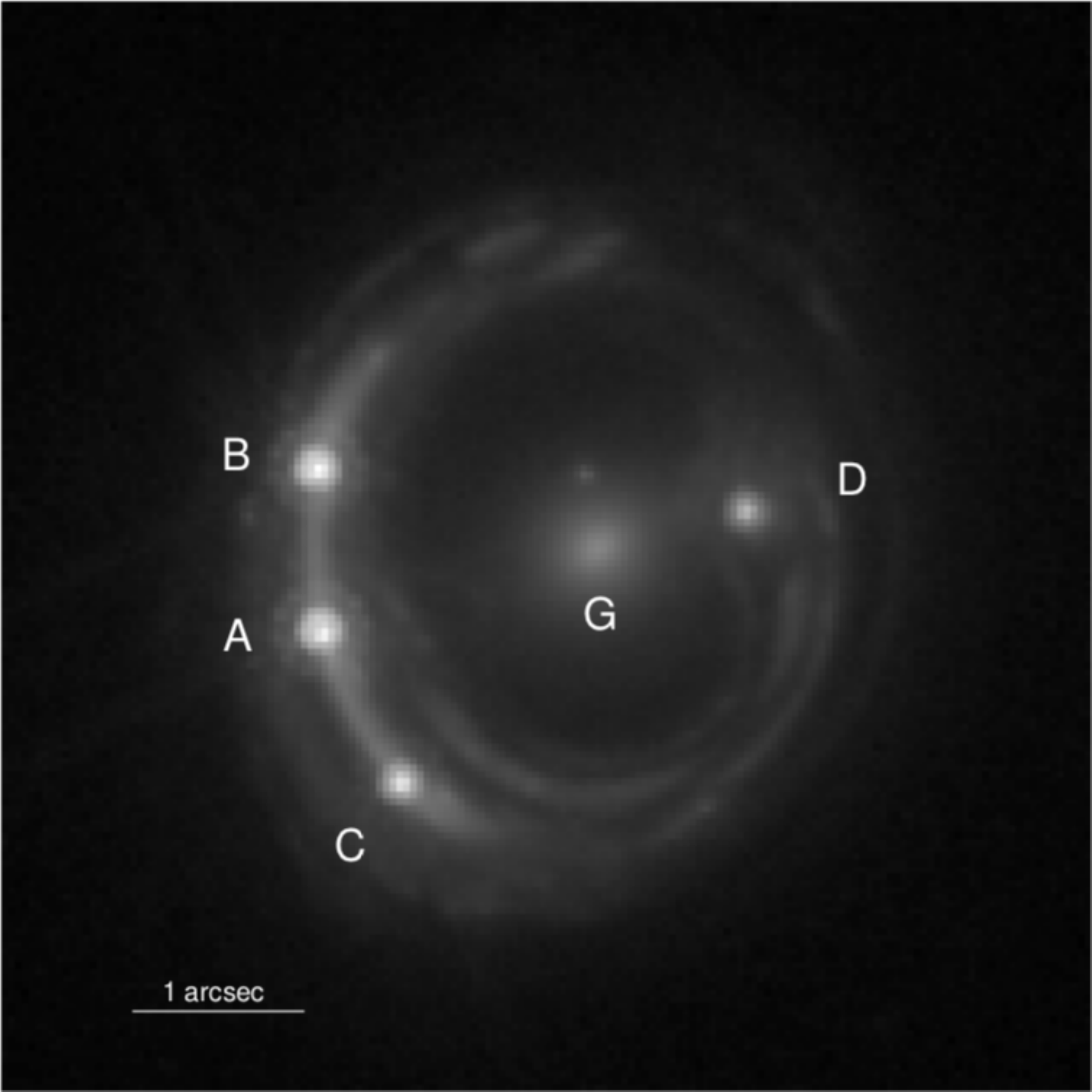}
  \caption{HST ACS image of \ourlens\ in the F814W filter. The lens galaxy G is surrounded by an Einstein ring and the four images of the quasar labelled with letters (A, B, C, D). The satellite in the northern part of the lens galaxy is not included in our mock data. The image resolution is $0.05''$/pixel}
  \label{fig:hst1131}
\end{figure}

\ourlens\ is a gravitational lensing system discovered by \citet{Sluse2003}. This system is composed of four multiple images of a distant quasar and its host galaxy \citep[$\zs=0.654$][]{Sluse2007}, lensed by an intervening giant elliptical galaxy at redshift $\zd=0.295$ \citep{Sluse2003,Sluse2007}, as shown in Figure \ref{fig:hst1131}. The quadruply imaged quasar is surrounded by an Einstein ring of $\sim$$3''$ diameter.
 This very peculiar system was used for studies on quasars and the region around black holes \citep[e.g.][]{Dai2010}, on time-delay cosmography \citep[e.g.][]{Morgan2006,Suyu2013, Suyu2014, Birrer2016,Chen2019}, on dark matter substructures \citep[e.g.][among others]{Birrer2017}, on planet searches with microlensing \citep[e.g.][]{Dai&Guerras2018}, and on black hole and galaxy co-evolution \citep[e.g.,][]{Ding2017}.
 Thanks to the large Einstein ring and the high magnification, the lensed quasar host promises to be an interesting test bed for obtaining IFU stellar kinematic data with future IFU instruments such as the {\rm Near Infrared Spectrograph} (NIRSpec) \citep[][]{Bagnasco2007,Birkmann2016} on the {\it James Webb Space Telescope} ({\it JWST}).
We estimate, through the Exposure Time Calculator\footnote{https://jwst.etc.stsci.edu} (ETC), that we would need an on-source integration time of $\rm \sim 6.5$ hours to obtain a signal-to-noise of $\sim$$11$ in the brightest pixel of the lensed arcs with NIRSpec. Through binning, we can easily increase the signal-to-noise ratio to 20, which is deemed necessary to properly extract the kinematic information from the data \citep{Barroso2017}. This shows how measuring spatially resolved stellar kinematics of galaxies in 2D at $\rm z\gtrsim 0.6$ would soon become feasible.

\subsection{\ourlens\ simulated data}
\label{subsec:SimData}

We construct mock lensing and kinematic data for \ourlens\ starting from the best-fit lens mass model obtained by fitting both the quasar image positions as point sources as well as the full Einstein ring as an extended source \citep{Yildirim2019} to the {\it Hubble Space Telescope} ($HST$) {\rm Advanced Camera for Surveys} (ACS) image of \ourlens\ in the F814W filter. The model of the lens mass distribution is a power law with an external shear (parametrised as described in Section \ref{subsec:MassLightProfiles}). The power law profile was already proven to be a good fit for this lensing system in previous studies \citep[among others]{ Suyu2013, Suyu2014} and, in general, power law profiles are a good representation of galaxies' mass distributions according to studies based on X-ray observations \citep{Humphrey&Buote2010} and the Sloan Lens ACS survey \citep[e.g.][]{KoopmansEtal06,Koopmans2009,Gavazzi2007,Auger2010,Barnabe2011} and from studies on lens potential corrections \citep{SuyuEtal09}. The best-fit parameters of the lens mass model for simulating our data are presented in Table \ref{tab:bestfitlens}. To probe dependencies on the parametrisation of the lens mass model we also explore a composite mass distribution of baryons and dark matter in Section \ref{subsubsec:Compositemodels}.

\begin{table*}
  \caption{Input lens mass parameters for \ourlens\ obtained by fitting both the quasar image positions as point sources as well as the full Einstein ring as an extended source \citep{Yildirim2019}. The lens mass distribution is constituted by a power law profile (SPEMD) plus an external shear, whose parameters are, respectively, the centroid position $x_{\rm c}^{\rm spemd}$,$y_{\rm c}^{\rm spemd}$, the axis ratio $q^{\rm spemd}$, the position angle $\theta^{\rm spemd}$, the Einstein radius for source at redshift infinity $\theta_{\rm E,\infty}^{\rm spemd}$, the core radius $r_{\rm core}^{\rm spemd}$ and the slope $\gamma^{\rm spemd}$ for the power law, the external shear strength for source at redshift infinity $\gamma_{\rm ext, \infty}$ and its orientation $\phi_{\rm ext}$.
  }
  \label{tab:bestfitlens}
  \begin{center}
  \begin{tabular}{lcccccccc}
    \hline
    \\
    $x_{\rm c}^{\rm spemd}$    & $y_{\rm c}^{\rm spemd}$    & $q^{\rm spemd}$ & $\theta^{\rm spemd}$ & $\theta_{\rm E,\infty}^{\rm spemd}$ & $r_{\rm core}^{\rm spemd}$ & $\gamma^{\rm spemd}$ & $\gamma_{\rm ext, \infty}$ & $\phi_{\rm ext}$ \\
    \\
    $['']$    & $['']$    &  & $[radians]$ & $['']$ & $['']$ &  & $['']$ &$[radians]$\\
    \\
    \hline\hline
    \\
    4.41 & 3.99 & 0.81 & 3.67 & 3.74 & $10^{-4}$ & 0.47 & 0.1 & 1.60\\ 
    \\
    \hline
  \end{tabular}
  \end{center}
%  Notes -- The top (bottom) half gives S\'{e}rsic profile parameters
%  from GALFIT for a two (single) component primary lens.  
%  Columns 1 and 2 denote the coordinate of the
%  centroid, column 3 is the effective radius, column 4 is the S\'{e}rsic
%  index, column 5 is the axis ratio, and column 6 is the position angle.

\end{table*}

\subsubsection{Mock lensing data}
\label{subsubsec:mocklensdata}
\begin{figure}
  \includegraphics[trim=0 0 0 0, clip,width=1.\columnwidth]{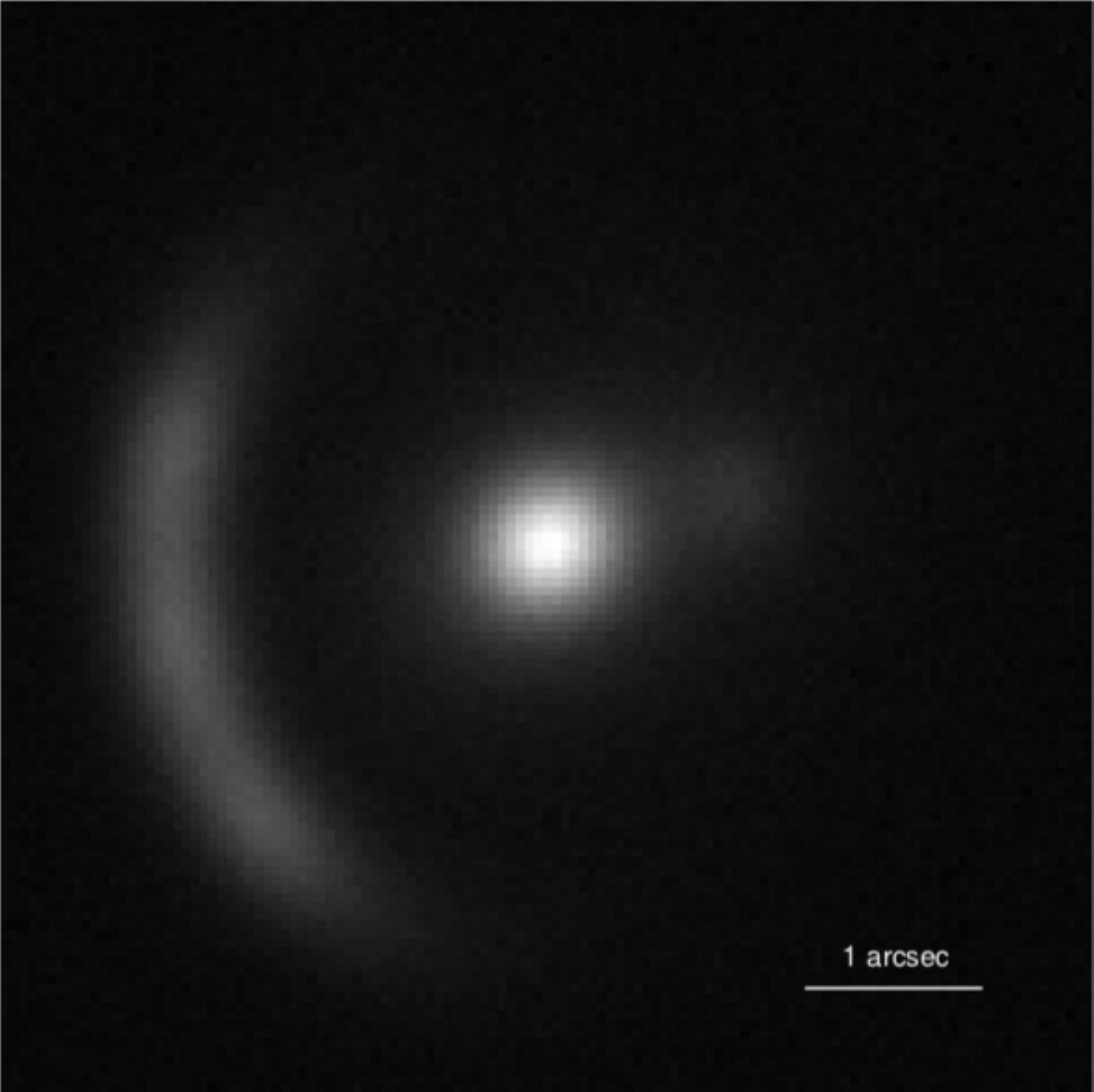}
  \caption{Mock surface brightness image of \ourlens\ obtained, as explained in Section \ref{subsubsec:mocklensdata}, using the parameters in Table \ref{tab:bestfitlens} and \ref{tab:mocklight}. The lens galaxy at the centre is surrounded by an Einstein ring and the four images of the host galaxy (quasar excluded). The image has a resolution of $0.05''$/pixel.}
  \label{fig:mock_SB}
\end{figure}

Given the lens mass model described in Section \ref{subsec:SimData}, we predict the centroid position of the quasar host galaxy on the source plane using GLEE. We parametrise the source by a \sersic\ in order to obtain a smooth surface brightness distribution, that we will use for the dynamical analysis as well.  We centre the source profile on the source centroid obtained with GLEE, and assume an axis ratio, orientation, effective radius and a \sersic\ index. We lens the so obtained source surface brightness map to the lens plane. To simulate the lens light, we centre a \sersic\ profile on the centroid coordinates of the lens mass profile described in Section \ref{subsec:SimData}, and we also assume the lens light orientation to be the same as that of the lens mass. We convolve the model with a Gaussian PSF with standard deviation of 4 pixels, which we estimated from measuring the FWHM of different stars in the field. 
Finally, we adjust the amplitudes of both light profiles, to obtain a surface brightness image that mimics the {\it HST} observations both in terms of the surface brightness distribution and the S/N. 
Since we mainly focus on the host galaxy's arcs, we have excluded the quasar's light from our model. The resulting surface brightness model is shown in Figure \ref{fig:mock_SB} and the input parameters are shown in Table \ref{tab:mocklight}. The data image has a total of $14641$ pixels.

\begin{table*}
  \caption{Input \sersic\ light parameters of the mock lens and source galaxies of \ourlens. The \sersic\ profile parameters are the centroid position $x_{\rm c}$ and $y_{\rm c}$, the axis ratio $q_{\rm s}$, the position angle $\theta_{\rm s}$, the \sersic\ amplitude $I_{\rm e,s}$, the effective radius $R_{\rm eff,s}$ and the \sersic\ index $n_{\rm s}$.   }
  \label{tab:mocklight}
  \begin{center}
  \begin{tabular}{lccccccc}
    \hline
    \\
    $ $ & $x_{\rm c}$    & $y_{\rm c}$    & $q_{\rm s}$ & $\theta_{\rm s}$ & $I_{\rm e,s}$ & $R_{\rm eff,s}$ & $n_{\rm s}$ \\
    \\
        & $['']$    & $['']$    &  & $[radians]$ & $[counts]$ & $['']$ &  \\
    \\
    \hline\hline
    \\
    Lens & 4.41 & 3.99 & 0.77 & 3.67 & 965. & 0.9 & 4 \\ 
    \\
    Source & 4.33 & 3.51 & 0.85 & 0.0 & 180.0 & 0.5 & 4 \\ 
    \\
    \hline
  \end{tabular}
  \end{center}
%  Notes -- The top (bottom) half gives S\'{e}rsic profile parameters
%  from GALFIT for a two (single) component primary lens.  
%  Columns 1 and 2 denote the coordinate of the
%  centroid, column 3 is the effective radius, column 4 is the S\'{e}rsic
%  index, column 5 is the axis ratio, and column 6 is the position angle.

\end{table*}

\subsubsection{Mock kinematic data}
\label{subsubsec:mockkindata}
\begin{figure}
  \includegraphics[trim=10 0 10 10, clip,width=\columnwidth]{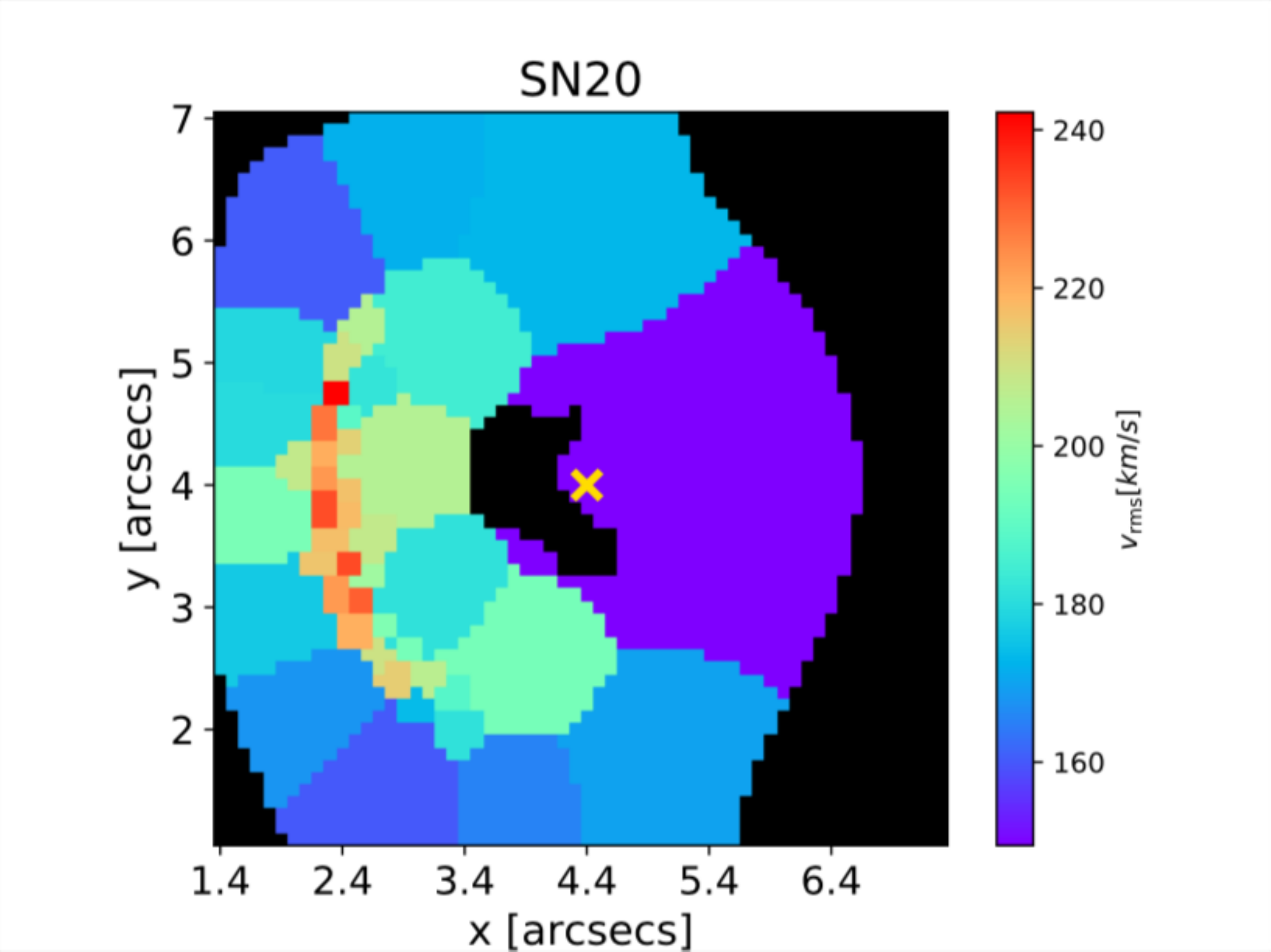}
  \caption{Mock binned lensed kinematic data of the source galaxy of \ourlens. We scale the signal of the source galaxy to obtain a signal-to-noise ratio in the brightest pixel of 15, to mimic a realistic scenario (see Section \ref{subsubsec:mockkindata} for further details). The data are binned to obtain a signal-to-noise in each bin of 20. We obtain a total of 45 bins. The pixel resolution of the kinematic map is $0.1''$. The masked black region corresponds to regions in the image plane which have no correspondence on the source plane. The yellow cross marks the lens centroid position.}
  \label{fig:mock_kin_SN20}
\end{figure}

\begin{figure}
  \includegraphics[trim=10 0 10 10, clip,width=\columnwidth]{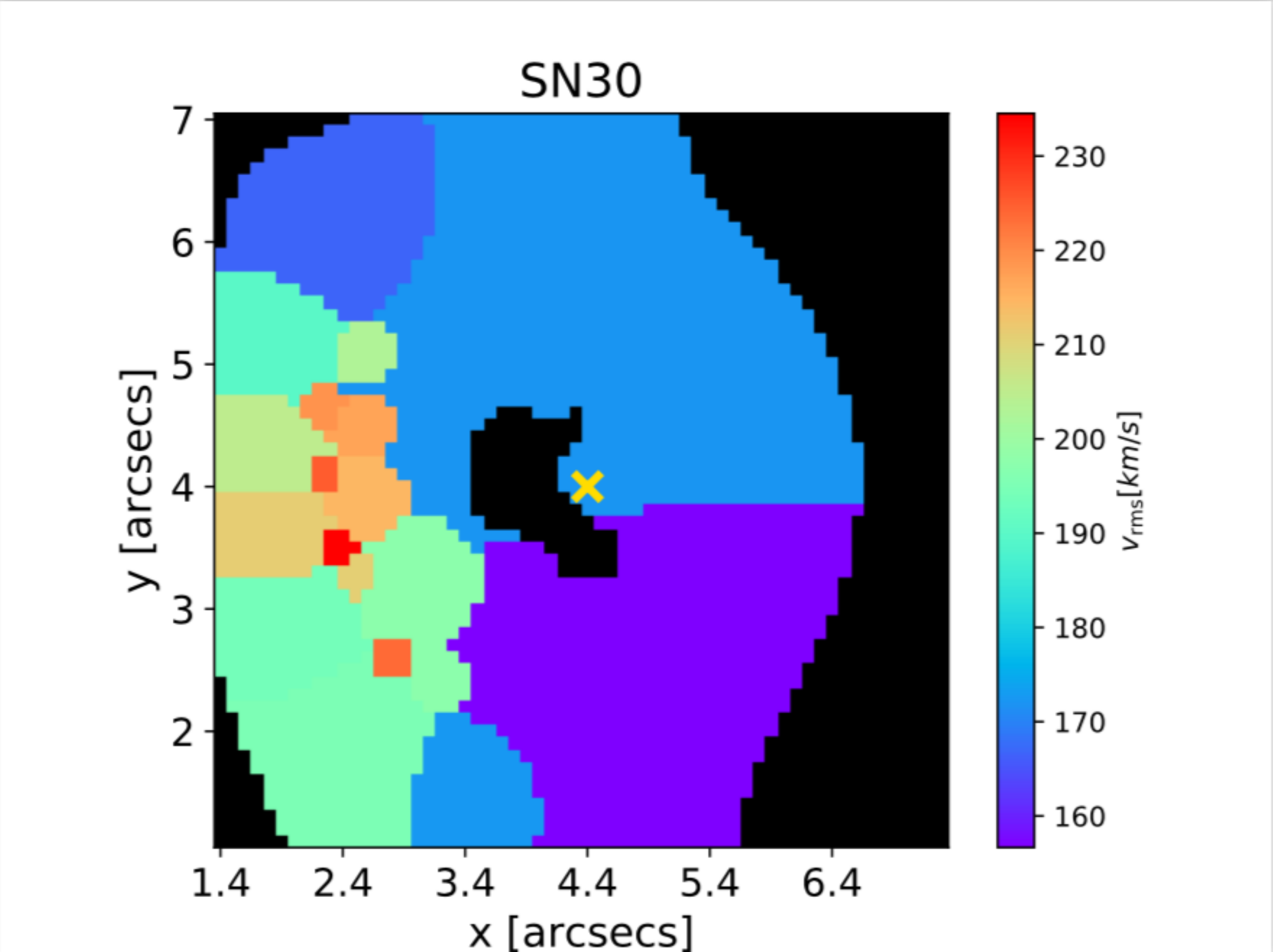}
  \caption{Mock lensed kinematic data of the source galaxy of \ourlens. We scale the signal of the source galaxy to obtain a signal-to-noise ratio in the brightest pixel of 15, to mimic a realistic scenario (see Section \ref{subsubsec:mockkindata} for further details). The data are binned to obtain a signal-to-noise in each bin of 30. We obtain a total of 19 bins. The pixel resolution of the kinematic map is $0.1''$. The masked black region corresponds to regions in the image plane which have no correspondence on the source plane.  The yellow cross marks the lens centroid position.}
  \label{fig:mock_kin_SN30}
\end{figure}

Once we have produced a mock lensing image to use as constraints, we produce mock kinematic data for the lensed arcs. To this end, we have to assume a mass profile for the source, in addition to the source (\sersic) light profile described in Section \ref{subsubsec:mocklensdata}. We adopt a singular pseudoisothermal elliptic mass distribution (PIEMD).  
We use a single profile for the combined dark and baryonic matter components given that the limited kinematic data in the arcs is most likely insufficient to effectively break the degeneracies between the two components. The parameters of this PIEMD profile are given in Table \ref{tab:mockmass}. We adopt the same centroid and same inclination for both the light and mass profiles of the source. We obtain the Einstein radius of the source mass distribution using the equation
\be
\label{eq:sigma}
\theta_{\rm E,\infty} = \frac{4 \pi \sigma_{\rm v}^{2}}{c^2},
\ee
where $\sigma_v$ is the mean velocity dispersion that we assume to be $200\ \rm km/s$ for the source galaxy, a sensible estimate for a galaxy at redshift $\zs=0.654$ hosting a black hole, whose mass we fixed to around $ 10^{8} M_{\odot}$ \citep{Dai2010}. Since we are studying the properties of the source, which does not act itself as a lens to any background galaxy, we compute all the quantities for an ``artificial'' source at redshift infinity, namely we fix the $\frac{D_{\rm ds}}{D_{\rm s}}=1$, where $D_{\rm ds}$ is the distance from our source at redshift $\zs=0.654$ to the ``artificial'' source at redshift infinity, and $D_{\rm s}$ is the distance to the ``artificial'' source. We show the source's mass parameters in Table \ref{tab:mockmass}.\\
\begin{table*}
  \caption{Input source mass parameters of the mock \ourlens. The source mass distribution is constituted by a pseudoisothermal elliptical mass distribution (PIEMD), whose parameters are the centroid position $x_{\rm c}^{\rm piemd}$,$y_{\rm c}^{\rm piemd}$, the axis ratio $q^{\rm piemd}$, the position angle $\theta^{\rm piemd}$, the Einstein radius for source at redshift infinity $\theta_{\rm E,\infty}^{\rm piemd}$ and the core radius $r_{\rm core}^{\rm piemd}$. We also show the values of the anisotropy $\beta$ and inclination $i$ we use to mock the kinematic data.
  }
  \label{tab:mockmass}
  \begin{center}
  \begin{tabular}{lccccccc}
    \hline
    \\
    $x_{\rm c}^{\rm piemd}$    & $y_{\rm c}^{\rm piemd}$    & $q^{\rm piemd}$ & $\theta^{\rm piemd}$ & $\theta_{\rm E,\infty}^{\rm piemd}$ & $r_{\rm core}^{\rm piemd}$ & $\beta$ & $i$\\
    \\
    $['']$    & $['']$    &  & $[radians]$ & $['']$ & $['']$ &  & $[radians]$\\
    \\
    \hline\hline
    \\
    4.33 & 3.51 & 0.90 & 0.0 & 1.15 & $5 \times 10^{-4}$ & $-0.15$ & 0.1 \\ 
    \\
    \hline
  \end{tabular}
  \end{center}
%  Notes -- The top (bottom) half gives S\'{e}rsic profile parameters
%  from GALFIT for a two (single) component primary lens.  
%  Columns 1 and 2 denote the coordinate of the
%  centroid, column 3 is the effective radius, column 4 is the S\'{e}rsic
%  index, column 5 is the axis ratio, and column 6 is the position angle.

\end{table*}
With the aforementioned parametrised profiles for the source mass and light, we produce an MGE model of the source light and mass on the source plane. We use the JAM routine to produce a model of the $\varv_{\rm rms}$ of the source galaxy on the source plane, and lens the kinematic map to the image plane. We then convolve with a gaussian PSF with zero mean and standard deviation of $0.1''$. Our final mock 2D kinematic map has a resolution of $0.1''$, to resemble the $JWST$ NIRSpec IFU instrument. As already mentioned in the introduction to this Section, we estimated the signal-to-noise ratio in the arcs to be $\sim$$11$ from 6.5 hours of observations with $JWST$ NIRSpec according to the ETC. 
In order to obtain a more favourable balance between S/N and spatial resolution of our source kinematic data, while keeping the total integration time still reasonable, we scale the surface brightness of the source light profile such that the signal-to-noise ratio of the brightest pixel on the image plane becomes $15$. To do that, we multiply the intensity by a factor $Q$ such that 
\be
\label{eq:err_dyn}
\left( \frac{S}{N} \right)_{\rm bp}=   \frac{I_{\rm bp} Q}{\sqrt{ \sigma_{\rm back}^{2} + I_{\rm bp} Q}}=15,
\ee
where `bp' denotes brightest pixel. We solve equation \ref{eq:err_dyn} to obtain the $Q$ factor and then scale the intensity of each pixel $I_{i}$ and the Poisson noise component of each pixel by multiplying them with the $Q$ factor. We therefore obtain a new rescaled intensity map with a signal-to-noise of 15 in the brightest pixel.\\
We use Voronoi binning as described in \cite{Cappellari&Copin2003}, an adaptive spatial binning technique, to obtain the signal-to-noise ratio per bin of $\sim$20 (labelled SN20) and $\sim$30 (labelled SN30), necessary to obtain reliable measurements of the stellar kinematics from the spectroscopic data. The binned kinematic map is illustrated in Figure \ref{fig:mock_kin_SN20} and \ref{fig:mock_kin_SN30} for the cases of SN20 and SN30, respectively. For simplicity, we assume that all bins contribute equally as constraints throughout the analysis. Realistically, however, the bins further away from the arc might not only have a lower constraining capacity than those closer to the arc, but it will be highly difficult to extract valuable kinematic information from the spectra, given the low S/N spaxels in these regions and the massive binning that would be needed to push the S/N above the threshold of 20 and 30, respectively. Therefore, before proceeding further with this assumption, we tested the case in which those bins have an artificially higher error on their kinematic moments, which effectively discards them from the fit. Since we noticed no significant discrepancy in the final parameter constraints, we assume all the bins to contribute equally throughout the analysis.

\subsection{Mass Models of Mock Data}
\label{subsec:models}
Once we have simulated a set of mock lensing and kinematic data, we perform different tests to assess how well we are able to recover the lens and source parameters. We sample the parameter space by varying the lens and source mass and light distributions. To assess the constraining power of the combination of the lensing and kinematic data, we model all the parameters using both the lensing and dynamics data (labelled LD) and using the lensing data only (labelled L). We also model only the lensing parameters (i.e. fixing the source mass) by using only the lensing data as constraints. We test this for both the SN20 and SN30 cases, to assess the improvement we obtain with a higher amount of kinematic constraints (model SN20 has more than double the number of bins than model SN30). In addition, we explore a model that has the source's mass distribution following its light (scaled by a M/L), which we denote as a mass-follows-light model (MFL), to assess systematic errors associated with the source mass parameterisation. Finally, we consider a different parametrization of the lens mass distribution to determine the impact of imperfect lens mass model on the inference of the source properties.

\subsubsection{LD: Lensing and Dynamical models}
\label{subsubsec:LDmodels}
Our first test consists of remodelling all the source's and lens' mass and light parameters using as constraints both the lensing and kinematic data, to check if GLaD is able to consistently recover the input parameters. We use both the kinematic data with SN20 and with SN30 (see Section \ref{subsubsec:mockkindata}). As shown in Table \ref{tab:ModelPara}, for both SN20 and SN30 most of the parameters are recovered within the $1\sigma$ uncertainties, and all are recovered within the $2\sigma$, as shown in Figure \ref{fig:cornerplot_LD}; both of the models have $\chi^2_{\rm red}\sim 1$. 
We note a tight anticorrelation between the \sersic\ amplitude $I_{\rm e,s}$ and the \sersic\ index $n$ of the source light, and between the source mass inclination and position angle. We also find a tight anticorrelation between the \sersic\ amplitude $I_{\rm e,s}$ and the effective radius of the lens light profile, and between the lens Einstein radius and the slope of the power-law profile. The source mass parameters do not show correlation with other parameters, apart from a mild anticorrelation between the source axis ratio and Einstein radius. Moreover, the inclination is not well constrained. We do not find a significant improvement on the lens galaxy constraints when using different kinematic data quality (SN20 and SN30), probably because the amount of kinematic data points (19 in the case of SN30 and 45 in the case of SN20) is still subdominant with respect to the lensing constraints (14641 pixels). However, we note a factor $\sim1.1-1.2$ improvement on the source parameters such as \sersic\ amplitude $I_{\rm e,s}$, effective radius $R_{\rm eff,s}$, \sersic\ index $n_{\rm s}$, source's mass axis ratio $q^{\rm piemd}_{\rm s}$ and Einstein radius $\theta^{\rm piemd}_{\rm E,s}$ in the SN20 case.
We expect this improvement to increase with higher signal-to-noise per pixel on the kinematic data, and a consequently higher amount of bins.
Therefore, for a given S/N in the kinematic data, it is preferable to spatially resolve as much as possible at the expense of a lower S/N per bin, provided that each bin has sufficient S/N to yield accurate $\varv_{\rm rms}$ measurements.

\subsubsection{L: Lensing only models}
\label{subsubsec:Lmodels}
\begin{figure*}
  \includegraphics[clip,width=1.\textwidth]{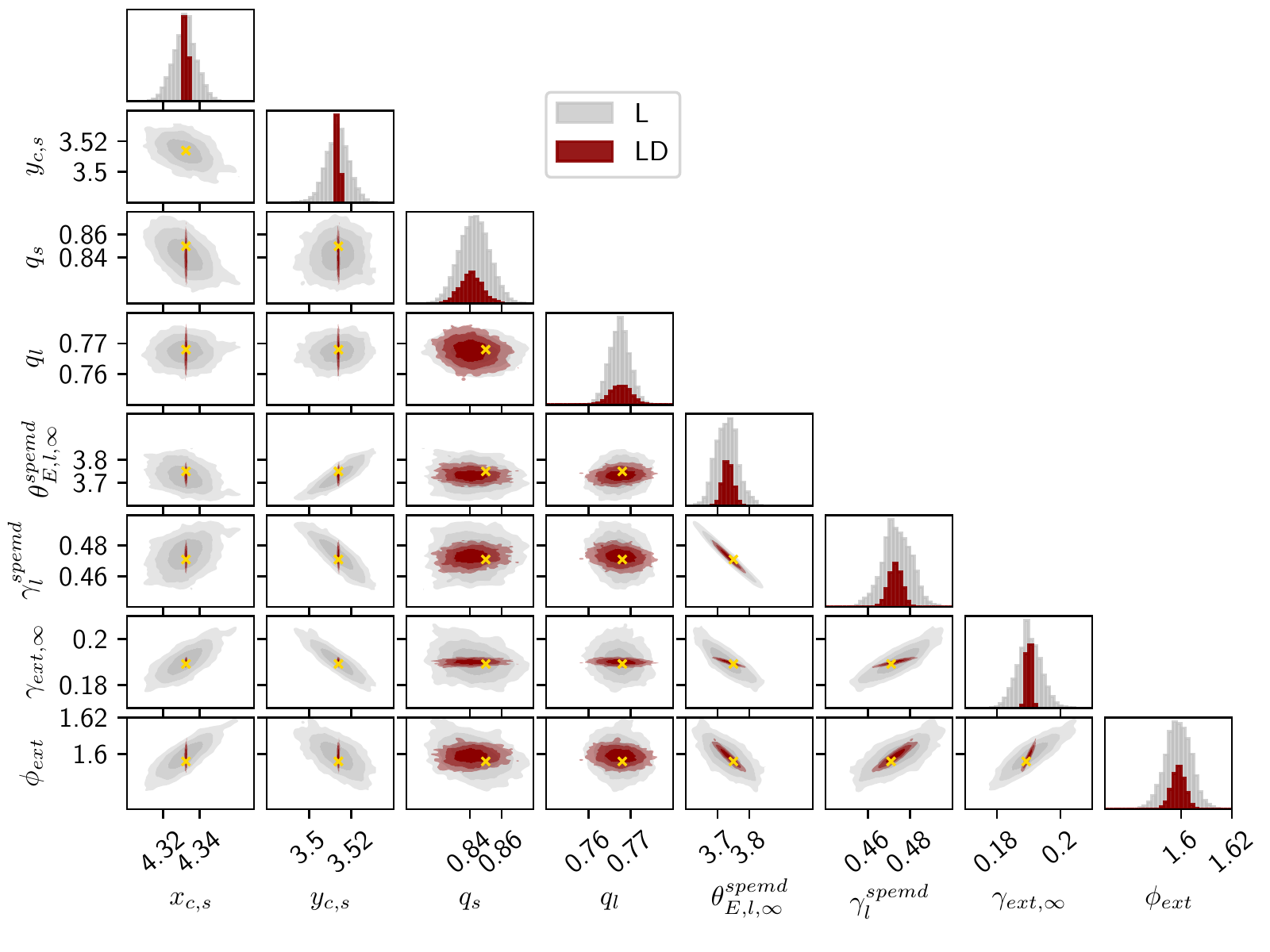}
  \caption{Joint 2D posterior probability distribution for the parameters of the power-law mock (presented in Section \ref{subsec:SimData}), with kinematic data having a signal-to-noise of 15 in the brightest pixel and binned to have a signal-to-noise of 30 in each bin. The different contours in the 2D plots indicate, respectively, the $1\sigma, 2\sigma$ and $3\sigma$ CIs. Parameters shown are those where the improvement on the constraints coming from the combination of lensing and dynamics (LD in red contours) is more prominent as compared to lensing only (L in grey contours). These parameters are the source centroid $x_{\rm c,s}, y_{\rm c,s}$, the anisotropy $\beta$, the source and lens light axis ratio $q_{\rm s}, q_{\rm l}$, the lens Einstein radius $\theta^{\rm spemd}_{\rm E,\infty,l}$ and slope $\gamma^{\rm spemd}_{\rm l}$, and the shear parameters $\gamma_{\rm ext, \infty}$ and $\phi_{\rm ext}$. In the diagonal are shown the 1D histograms of the corresponding parameter on the x-axis.}
  \label{fig:cornerplot_LD}
\end{figure*}

\begin{figure*}
  \includegraphics[clip,width=1.\textwidth]{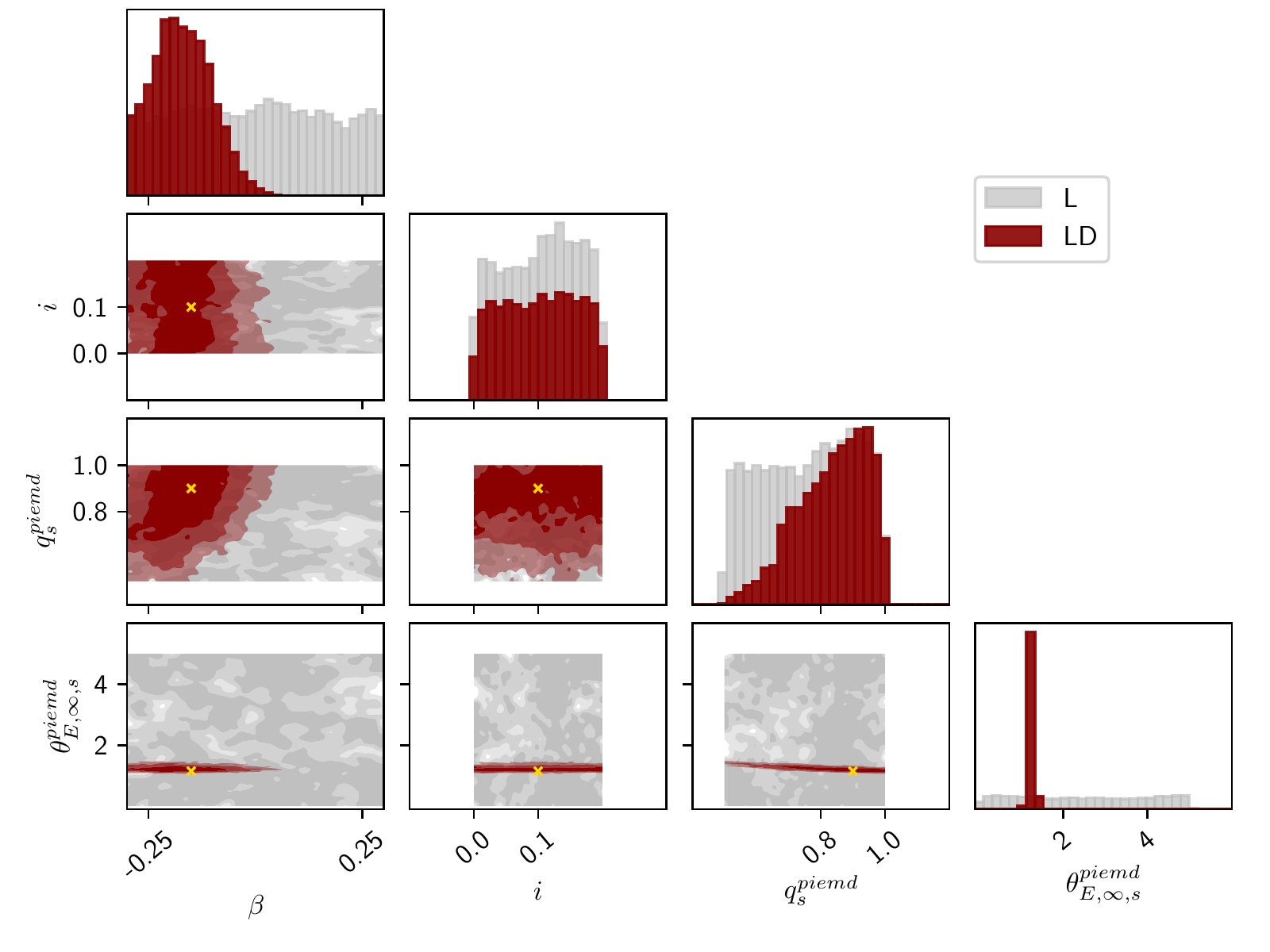}
  \caption{Joint 2D posterior probability distribution for the source mass parameters of the power-law mock data presented in Section \ref{subsec:SimData}, with kinematic data having a signal-to-noise of 15 in the brightest pixel and binned to have a signal-to-noise of 30 in each bin. The different contours in the 2D plots indicate the different $\sigma$ CIs. Parameters shown are those where the improvement on the constraints coming from the addition of lensing and dynamics (LD in red contours) is more prominent as compared to lensing only (L in grey contours). These parameters are the anisotropy $\beta$, the inclination $i$, the source mass axis ratio $q^{\rm piemd}_{\rm s}$, the source Einstein radius $\theta^{\rm piemd}_{\rm E,\infty, s}$. In the diagonal are shown the 1D histograms of the corresponding parameter on the x-axis.}
  \label{fig:cornerplot_LD_sourcemass}
\end{figure*}

In the second test we perform, we include only the lensing surface brightness data as constraints, in order to assess the changes when compared to the joint lensing and dynamical modelling in Section \ref{subsubsec:LDmodels}.
As shown in Table \ref{tab:ModelPara} and in Figure \ref{fig:cornerplot_LD}, parameters like the source mass and light centroid, the source \sersic\ amplitude $I_{\rm e,s}$ and effective radius are better constrained (up to an order of magnitude tighter) in the combined lensing and dynamics case. Indeed, the dynamical analysis makes direct use of these quantities to predict the model of the lensed source kinematic map. Moreover, we note that also the lens amplitude, lens mass axis ratio and Einstein radius are constrained better up to a factor of 3 in the combined lensing and dynamics case. This shows that actually the addition of the kinematic data will put tighter constraints on the lens parameters as well. We also find that the addition of the kinematic data puts tighter constraints on the shear parameters (amplitude and orientation), with respect to lensing only case. As expected, when trying to vary the source mass parameters not including the kinematics analysis, we find them to be virtually unconstrained, as shown in Table \ref{tab:ModelPara} and in Figure \ref{fig:cornerplot_LD_sourcemass}. Finally, we note a tight anticorrelation between the \sersic\ amplitude $I_{\rm e,s}$ and the \sersic\ index $n$ of the source light, and between the source mass inclination and position angle, as already seen in the LD case. For this model, the input parameters are mostly recovered within the $1\sigma$ uncertainties, and are all recovered within the $2\sigma$, with a $\chi^2_{\rm red}\sim 1$.
\subsubsection{MFL: Mass-follows-light models}
\label{subsubsec:MFLmodels}
\begin{figure*}
  \includegraphics[clip,width=1.\textwidth]{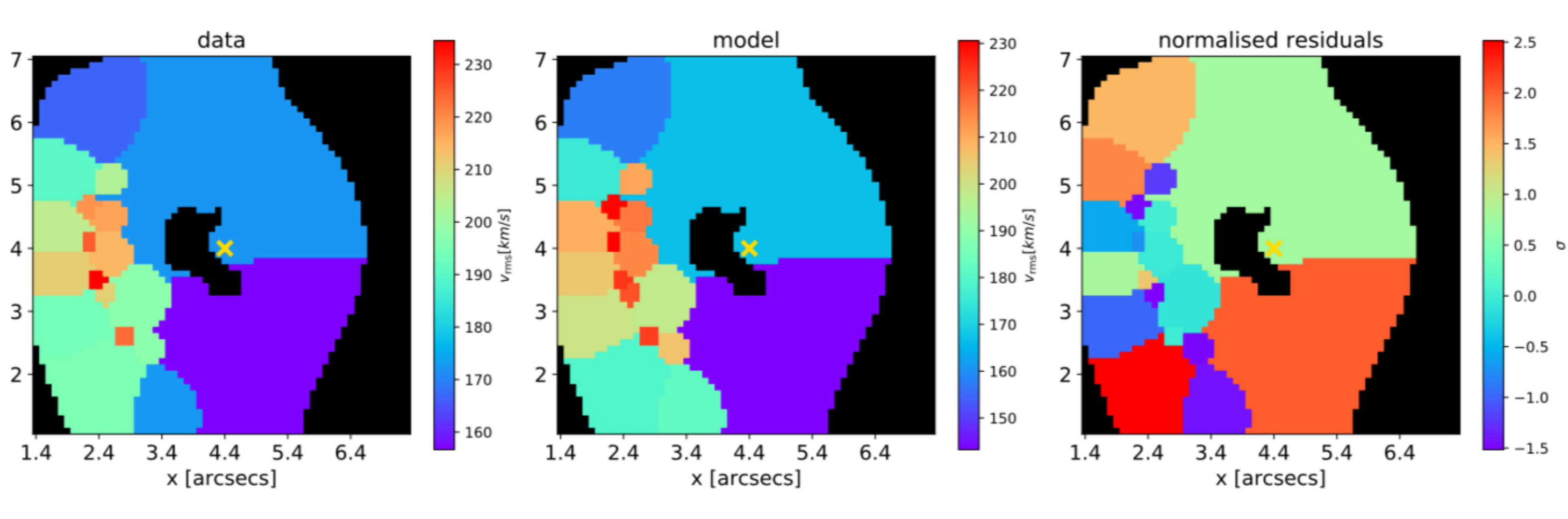}
  \caption{Mock lensed kinematic data of the source galaxy of \ourlens\ obtained using an isothermal mass profile (left panel), reconstructed kinematic map from our best-fit MFL model (central panel), and normalised residuals (right panel). The pixel resolution of the kinematic map is $0.1''$. The masked black region is the region in the image plane which has no correspondence on the source plane. The yellow cross marks the lens centroid position. If we compare to the data, we see that our model predicted source's $\varv_{\rm rms}$ is slightly underestimated.}
  \label{fig:Dyn_MFL}
\end{figure*}

\begin{figure*}
  \includegraphics[trim=100 0 100 0,clip,width=1.\textwidth]{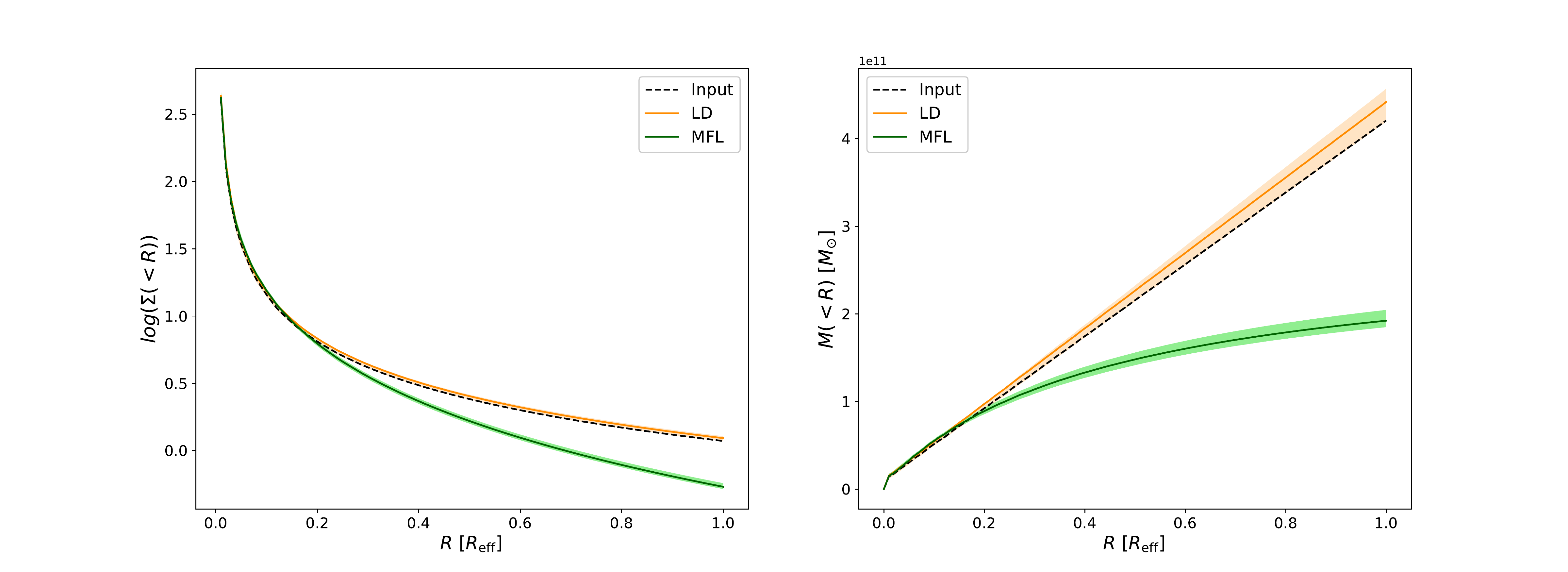}
  \caption{Predicted averaged circularised surface mass density $\Sigma(<R)$ (left) and mass enclosed $\rm M(<R)$ (right) within one effective radius $R_{\rm eff}=0.5''$ ($3.5$ kpc) of the source galaxy, with the corresponding CI (shaded regions). Both the convergence and the enclosed mass of the source of the MFL model (see Section \ref{subsubsec:MFLmodels}) are underestimated in the outskirts, as compared to the input and the LD model. This is also shown in the $\varv_{\rm rms}$ value, which is slightly underestimated, as shown in Figure \ref{fig:Dyn_MFL}.}
  \label{fig:Dyn_MFL_kappa}
\end{figure*}

To test systematic modelling uncertainties, we model the source mass (originally simulated as an isothermal profile) using a different model. We use a mass-follows-light model, i.e., we scale the source light profile (\sersic) using a mass-to-light ratio, and we allow the mass-to-light ratio to vary together with the other parameters. We fit to the lensing and kinematic data (those binned to obtain a signal-to-noise of 30 in each bin, SN30), and we test how well this model is able to reproduce the original data. \\
We find that assuming a different source mass profile biases some of the model parameters, which are not recovered within the uncertainties, as shown in Table \ref{tab:ModelPara}. In this particular case, the MFL model cannot recover most of the source light, mass and dynamical parameters within the $1\sigma$ uncertainties, but most of them are recovered within the $2\sigma$. The only parameter which is overestimated by $\sim 4\sigma$ is the anisotropy. The source inclination, \sersic\ index, and light intensity seem to be more robust parameters and are less affected by systematics in this particular case, as they are mainly anchored by the lensing data. For the lens, we find that the parameters affected by systematics are the \sersic\ intensity and the mass Einstein radius, ellipticity and slope, which are all tightly correlated to each other, as already mentioned for the LD case in Section \ref{subsubsec:LDmodels}. Finally, the external shear parameters are not recovered within the $3\sigma$ uncertainties. This might be the cause of the shift of the source centroid prediction, which is sensitive to the shear strength value, and explain the misfit of the kinematic data. \\
This model has a slightly higher $\chi^2_{\rm red}$ than the other models ($\chi^2_{\rm red}\sim1.07$), showing that it does not fit as well to the data. In particular, the dynamics $\chi^2_{\rm dyn,red}$ is $\sim 1.63$, while the lensing $\chi^2_{\rm len,red}$ is $\sim 1.07$, which tells us that the kinematic data are poorly fitted as compared to the lensing data. Indeed, as shown in Figure \ref{fig:Dyn_MFL}, the $\varv_{\rm rms}$ is underestimated, especially in the bins around the arc region. The surface mass density $\Sigma(<R)$, shown in Figure \ref{fig:Dyn_MFL_kappa}, is in good agreement with the data for radii smaller than $0.2\ R_{\rm eff}$, and underestimated otherwise, by up to a factor 2. We find that the size of the region within $0.2\ R_{\rm eff}$ from the center of the source on the source plane maps exactly to the location of the brighest bins (those with high signal-to-noise) on the image plane. This indicates that our constraints are robust in the region of high signal-to-noise, and instead model dependent on the regions with low signal-to noise. Therefore, we conclude that the addition of dynamics can help better distinguish between models, whereas the lensing analysis alone provides no constraints on the source mass at all.
\subsubsection{Composite models}
\label{subsubsec:Compositemodels}
\begin{figure*}
  \includegraphics[trim=0 0 0 10,clip,width=1.\textwidth]{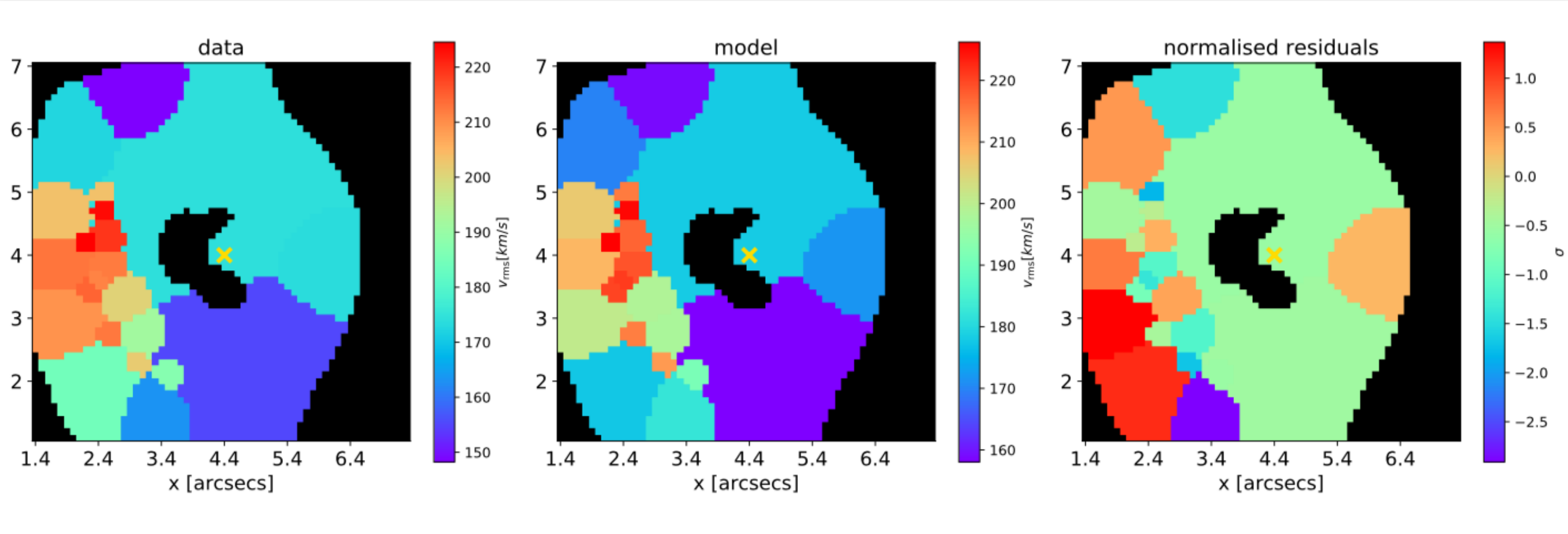}
  \caption{Mock lensed kinematic data of the source galaxy of \ourlens\ for the composite mass model (left panel), our prediction obtained by modelling the data with a power law profile (central panel) and the normalised residuals (right panel). The data are binned to obtain a signal-to-noise in each bin of 30, for a total of 22 bins. The pixel resolution of the kinematic maps is $0.1''$. The masked black region is the region in the image plane which has no correspondence on the source plane. The yellow cross marks the lens centroid position. }
  \label{fig:mock_kin_SN30_comp}
\end{figure*}

\begin{figure*}
  \includegraphics[trim=0 0 0 0,clip,width=1. \textwidth]{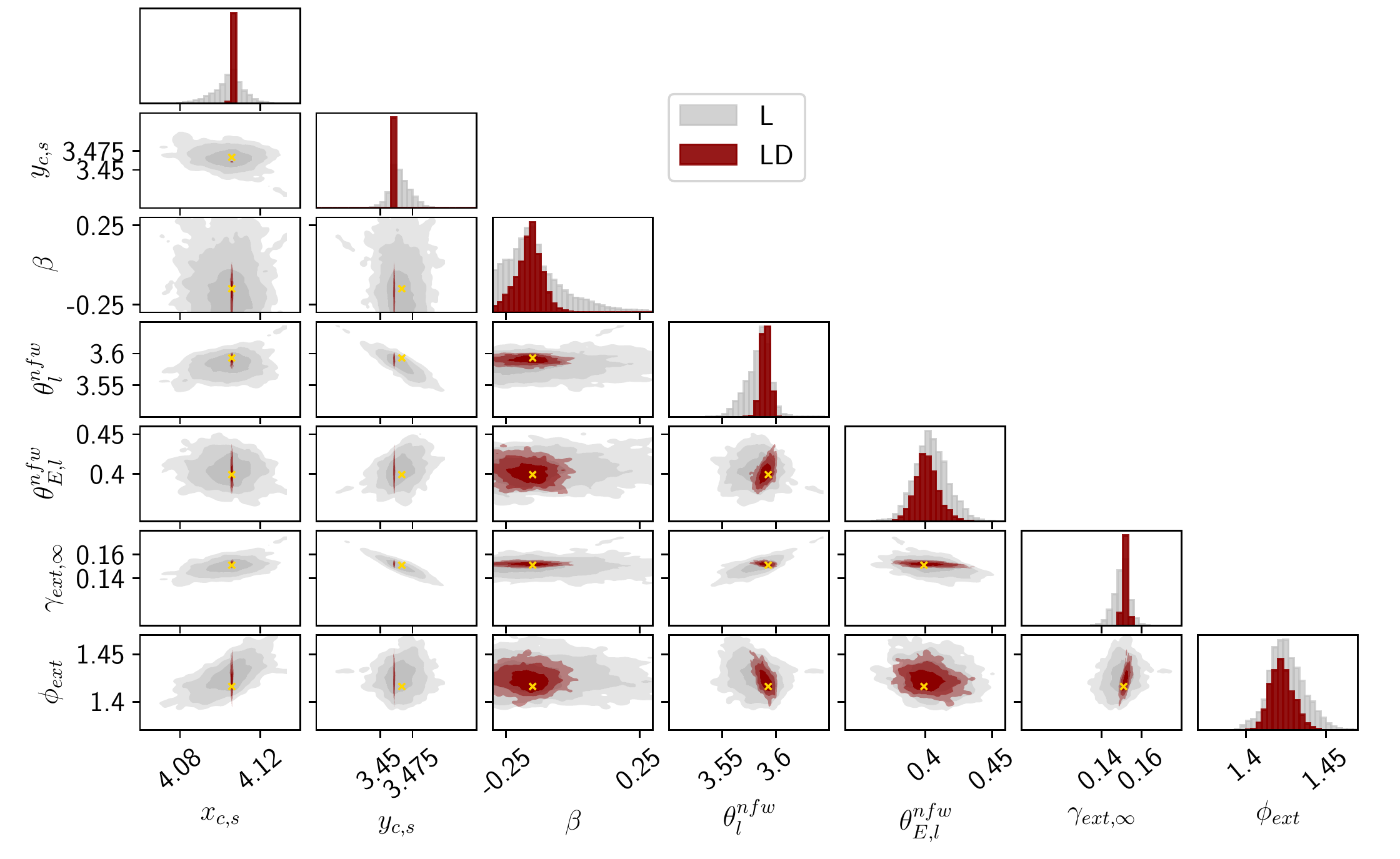}
  \caption{Joint 2D posterior probability distribution for the parameters of the composite mock (presented in Section \ref{subsubsec:Compositemodels}), with kinematic data having a signal-to-noise of 15 in the brightest pixel and binned to have a signal-to-noise of 30 in each bin (shown in Figure \ref{fig:mock_kin_SN30_comp}). The different contours in the 2D plots indicate, respectively, the $1\sigma, 2\sigma$ and $3\sigma$ ranges. Parameters shown are those where the improvement on the constraints coming from the combination of lensing and dynamics (LD in red contours) is more prominent as compared to lensing only (L in grey contours). These parameters are the source centroid $x_{\rm c,s}, y_{\rm c,s}$, the anisotropy $\beta$, the position angle of the lens dark matter profile $\theta^{\rm nfw}_{\rm l}$, its Einstein radius $\theta^{\rm nfw}_{\rm E,\infty,l}$, and the shear parameters $\gamma_{\rm ext, \infty}$ and $\phi_{\rm ext}$. In the diagonal are shown the 1D histograms of the corresponding parameter on the x-axis.
}
  \label{fig:cornerplot_LD_composite}
\end{figure*}

\begin{figure*}
  \includegraphics[trim=10 0 10 0, clip,width=1.\textwidth]{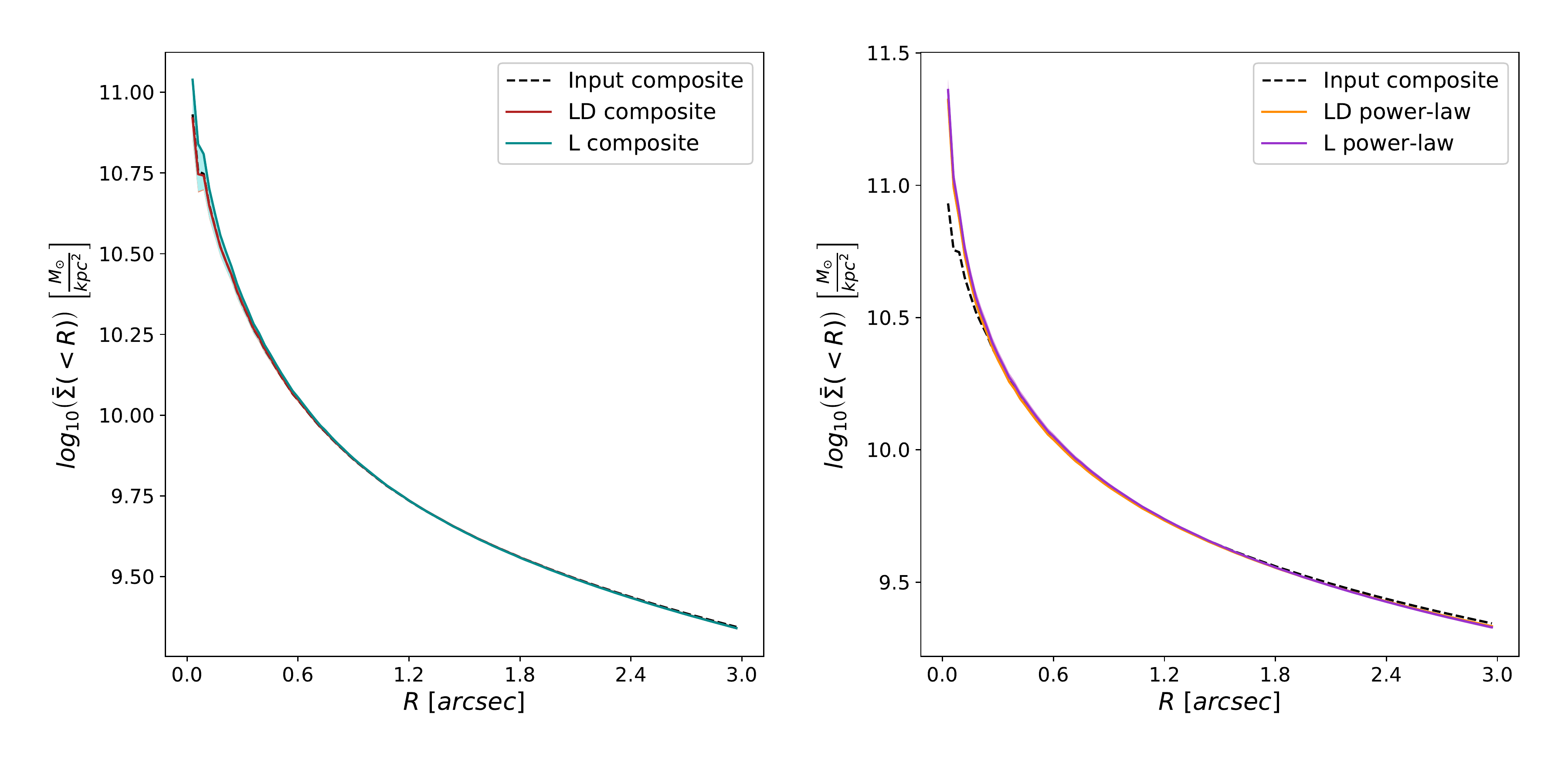}
  \caption{Average surface mass density $\bar{\Sigma}(R)$ for the LD composite and the L composite models (left panel), and for the power-law models LD power-law and L power-law (right panel) as compared to the input composite mock, plotted with the relative error bars. All models recover the input closely at the value of the Einstein radius ($\sim 1.6''$ for source at reshift $0.654$). On the left panel, the L composite model gets peakier towards the center, but is still able to recover the input within the error bars (light blue shaded region). This does not appear to be the case for the LD power-law and L power-law (right panel), which differ from the input for $R<0.3''$. However, at larger radii, we cannot easily distinguish between power-law and composite models from the average surface mass density.}
  \label{fig:ASMD_comp}
\end{figure*}

As a final test, we assess the improvement of the constraints on the lens mass model, particularly focussing on the dark matter component. To do so, we re-simulate both the lensing and kinematics data, as done in Section \ref{subsec:SimData}, assuming a composite mass model for the lens. In particular, to represent the baryonic mass we use a chameleon profile (discussed in Section \ref{subsec:MassLightProfiles}) that we scale with a mass-to-light ratio, and for the dark matter we assume a NFW profile. We still assume an external shear component. For this model, the lens \sersic\ light and the lens mass are decoupled. We show the parameters of this model in Table \ref{tab:ModelParaComposite}. For the kinematic data, we imposed a signal-to-noise of 30 in each bin (SN30), which allows us to obtain 22 kinematic data points (shown in Figure \ref{fig:mock_kin_SN30_comp}). We remodel these new simulated data with both the full lensing and dynamics configuration and the lensing only configuration, to assess the improvement due to the addition of the kinematic constraints. Then we remodel the lens mass using a power-law model, to assess systematic uncertainties associated with the lens mass parameterisation.  As shown from Table \ref{tab:ModelParaComposite} and Figure \ref{fig:cornerplot_LD_composite}, we find improved constraints on the source centroid when we add the kinematic constraints, consistent with our findings for the power-law model. Moreover, the dark matter (NFW profile) orientation and Einstein radius are better constrained when including the kinematic data, compared to the lensing-only analysis. Both the LD composite and the L composite model have a $\chi_{\rm red}^2\sim 1.$ 
When testing systematics errors for this set of mock data with the composite model, i.e. when modelling the lens composite mass with a single power-law model, we find that the addition of the kinematic data allows to better discern between models. Indeed, as shown in Table \ref{tab:ModelParaComposite}, the LD power-law model has a comparable total $\chi_{\rm red}^2$($\chi_{\rm red}^2=1.03$) than the lensing only L power-law ($\chi_{\rm red}^2\sim 1.02$). However, the LD power-law has a higher misfit of the kinematic data $\chi_{\rm dyn,red}^2\sim 1.25$. Moreover, in the LD power-law we find a better constrained lens Einstein radius by a factor 2 as compared to the L power-law. This is consistent with the previous set of mock data. Finally, as already noted previously, the source parameters are constrained when including kinematic data, in both the input set-up and the systematic test. In general, systematic uncertainties, from profile mismatch, dominate statistical uncertainties. Encouragingly, most of the mass parameters of the sources are recovered within $1\sigma$ uncertainty in the LD power-law model, showing that the source mass distribution could still be inferred from the kinematics despite differences in the lens mass profile between power-law and composite profiles.\\
To show more in detail the differences among these models, we compare their average surface mass density\footnote{Average circularised, since we integrate all the mass distributions over circles.}, namely
\be
\label{eq:ASMD}
\bar{\Sigma} (<R) = \frac{\int_{0}^{R} \! \Sigma(R') 2 \pi R'  \, \mathrm{d}R' } {\pi R^2}.
\ee
where $\Sigma(R)$ is the surface mass density. The $\bar{\Sigma}(<R)$ for the input, LD composite and L composite, is shown in the left panel of Figure \ref{fig:ASMD_comp}, while the $\bar{\Sigma}(<R)$ for the LD power-law and L power-law models is shown in the right panel. From Figure \ref{fig:ASMD_comp} we see that the LD composite and L composite models are very consistent with the input in terms of average surface mass density. When we model with the power-law lens mass model, we still see a strong consistency with the input for the $\bar{\Sigma} (<R)$. We note mild inconsistencies at  $R< 0.3''$, where the LD power-law and L power-law models get peakier than the input. \\
Therefore we conclude that, for this system, the composite and power law models are not distinguishable in terms of $\bar{\Sigma} (<R)$ and of the lensing analysis, but are thanks to the use of the dynamics analysis, which shows the higher misfit when a different lensing model than the input is used.
\begin{table*}
\caption{Lens and source parameter constraints for the different configurations for our first set of mock data (presented in Section \ref{subsec:SimData}). The columns show the best-fitting values with the corresponding $1\sigma$ uncertainties for the joint lensing and dynamical models (LD) with different signal-to-noise ratios for the binning (SN20 and SN30), for the lensing-only model (L), and finally for models with a mass-follows-light profile for the source mass distribution (with SN30) to test for systematic uncertainties related to the source mass parameterisation.}
% title of Table
\label{tab:ModelPara}      % is used to refer this table in the text
\centering                          % used for centering table
\renewcommand{\arraystretch}{1.4}  % change the cell height (row spacing)
\begin{tabular}{llcccccc}        % centred columns (3 columns)
\hline                 % inserts double horizontal lines
\hline
Parameter description & Parameter  &  Input & Prior & LD (SN20) & LD (SN30) & L & MFL \\    % table heading 
\hline
\hline                        % inserts single horizontal line
%\hline
 & Source &  &    &  &   &   &\\
\hline
& light &  &    &  &   &   & \\
x centroid & $\phantom{ }x_{\rm c,s}$     $['']$                 &    $4.3325$ & (3.,8.)  & $4.3325_{-0.0003}^{+0.0003}$ & $4.3325_{-0.0002}^{+0.0002}$ & $4.332_{-0.007}^{+0.006}$  & $4.11_{-0.01}^{+0.01}$ \\ 
y centroid & $\phantom{ }y_{\rm c,s}$       $['']$               &   $3.5139$  &  (3.,8.) & $3.5138_{-0.0002}^{+0.0002}$ & $3.5139_{-0.0002}^{+0.0002}$ & $3.514_{-0.005}^{+0.005}$ & $3.594_{-0.005}^{+0.007}$ \\ 
axis ratio & $\phantom{ }q_{\rm s}$                      &$0.850$ &  (0.5,1.) & $0.842_{-0.007}^{+0.007}$& $0.844_{-0.008}^{+0.008}$ & $0.843_{-0.010}^{+0.009}$ & $0.91_{-0.02}^{+0.02}$ \\ 
position angle & $\phantom{ }\theta_{\rm s}$     $[radians]$                 & $0.00$ &  - & $-0.01_{-0.03}^{+0.02}$& $-0.02_{-0.03}^{+0.02}$ & $-0.02_{-0.03}^{+0.03}$ & $-0.91_{-0.04}^{+0.07}$ \\ 
intensity & $\phantom{ }I_{\rm e,s}$        $[counts]$              &$180$ & (0.,300.)  & $182._{-7.}^{+8.}$ & $182._{-8}^{+9.}$ & $184._{-9.}^{+11.}$ & $189._{-11.}^{+12.}$ \\ 
effective radius & $\phantom{ }R_{\rm eff,s}$                      &$0.500$ &  (0.2,2.) & $0.50_{-0.01}^{+0.01}$& $0.50_{-0.01}^{+0.01}$ & $0.50_{-0.02}^{+0.02}$ & $0.42_{-0.01}^{+0.02}$ \\ 
sersic index & $\phantom{ }n_{\rm s}$                      &$4.0$ &  (0.5,10.) & $3.98_{-0.06}^{+0.05}$ & $3.98_{-0.06}^{+0.06}$ & $3.96_{-0.07}^{+0.07}$ & $3.95_{-0.09}^{+0.07}$ \\ 
\hline
& mass &  &   &   &   &  &  \\
anisotropy & $\phantom{ }\beta$                      & $-0.15$ & (-0.3,0.3)  & $-0.16_{-0.07}^{+0.05}$ & $-0.11_{-0.05}^{+0.08}$ & $-0.1_{-0.2}^{+0.2}$ & $0.04_{-0.05}^{+0.05}$ \\ 
inclination & $\phantom{ }i$                      & $0.10$ & (0.,0.2)  &$0.10_{-0.06}^{+0.06}$ & $0.10_{-0.05}^{+0.06}$& $0.10_{-0.07}^{+0.07}$ & $0.11_{-0.07}^{+0.06}$ \\ 
mass-to-light ratio & $\phantom{ }\frac{M}{L}$                      & - & (1.,7.)  & - & - & - & $0.003_{-0.0001}^{+0.0001}$  \\ 
axis ratio & $\phantom{ }q^{\rm piemd}_{\rm s}$                      & $0.90$ &  (0.5,1.)  &$0.89_{-0.12}^{+0.05}$ & $0.85_{-0.18}^{+0.07}$ & $0.8_{-0.2}^{+0.1}$ & -\\
Einstein radius& $\phantom{ }\theta^{\rm piemd}_{\rm E, \infty,s}$    $['']$                 & $1.15$ &  (0.,5.) & $1.19_{-0.04}^{+0.07}$ &  $1.19_{-0.04}^{+0.08}$& $1.2_{-0.6}^{+2.1}$ & - \\
\hline
& Lens &  &   &   &   &  &  \\
\hline
& light &  &   &   &   &  & \\
axis ratio & $\phantom{ }q_{\rm l}$                      & $0.768$ & (0.2,1.)  & $0.768_{-0.002}^{+0.002}$ & $0.768_{-0.002}^{+0.002}$ & $0.767_{-0.003}^{+0.002}$ & $0.771_{-0.003}^{+0.003}$ \\ 
intensity & $\phantom{ }I_{\rm e,l}$                       & $965.$ &  (800.,1100.) & $959._{-21.}^{+15.}$ & $958._{-23.}^{+17.}$ & $951._{-26}^{+22}$ & $910._{-26.}^{+35.}$ \\ 
effective radius & $\phantom{ }R_{\rm eff,l}$                      & $0.900$ & (0.3,2.) & $0.904_{-0.008}^{+0.011}$ & $0.904_{-0.009}^{+0.012}$ & $0.91_{-0.01}^{+0.01}$ & $0.92_{-0.02}^{+0.02}$\\ 
Sersic index & $\phantom{ }n_{\rm l}$                      &$4.00$ & (0.5,10.)  & $4.01_{-0.03}^{+0.04}$ & $4.01_{-0.03}^{+0.04}$ & $4.03_{-0.04}^{+0.05}$ & $4.15_{-0.07}^{+0.05}$ \\ 
\hline
& mass &  &      &   &   &   &\\
x centroid & $\phantom{ }x^{\rm spemd}_{\rm c,l}$       $['']$               &  $4.4083$   & (4.,5.)  & $4.4085_{-0.0003}^{+0.0004}$ & $4.4086_{-0.0004}^{+0.0005}$ & $4.4087_{-0.0004}^{+0.0005}$ & $4.4062_{-0.0005}^{+0.0005}$ \\ 
y centroid & $\phantom{ }y^{\rm spemd}_{\rm c,l}$        $['']$            &  $3.9934$  & (3.5,5.5)  & $3.9935_{-0.0003}^{+0.0003}$ & $3.9934_{-0.0003}^{+0.0003}$ & $3.9935_{-0.0003}^{+0.0004}$ & $3.9940_{-0.0005}^{+0.0004}$\\ 
axis ratio & $\phantom{ }q^{\rm spemd}_{\rm l}$                      &$0.808$ & (0.2,1.)  & $0.807_{-0.002}^{+0.001}$ & $0.806_{-0.002}^{+0.002}$ & $0.806_{-0.004}^{+0.003}$ & $0.759_{-0.004}^{+0.013}$ \\ 
position angle & $\phantom{ }\theta^{\rm spemd}_{\rm l}$      $[radians]$                &$3.669$ & -  & $3.666_{-0.004}^{+0.003}$ & $3.665_{-0.004}^{+0.004}$ & $3.664_{-0.005}^{+0.005}$ & $3.597_{-0.005}^{+0.008}$ \\ 
Einstein radius & $\phantom{ }\theta^{\rm spemd}_{\rm E,\infty, l}$    $['']$                 & $3.75$ & -  & $3.74_{-0.02}^{+0.01}$ & $3.73_{-0.02}^{+0.02}$ & $3.73_{-0.04}^{+0.03}$ & $4.11_{-0.06}^{+0.11}$ \\ 
slope & $\phantom{ }\gamma^{\rm spemd}_{\rm l}$                      & $0.471$ & (0.2,0.8)  & $0.472_{-0.003}^{+0.003}$ & $0.473_{-0.003}^{+0.003}$ & $0.473_{-0.006}^{+0.007}$ & $0.407_{-0.014}^{+0.008}$\\ 
shear strength & $\phantom{ }\gamma_{\rm ext, \infty}$    $['']$                  &$0.1892$ & (0.,0.3)  & $0.1898_{-0.0006}^{+0.0007}$ & $0.1900_{-0.0007}^{+0.0007}$ & $0.190_{-0.003}^{+0.004}$ & $0.111_{-0.002}^{+0.003}$ \\ 
shear position angle & $\phantom{ }\phi_{\rm ext}$        $[radians]$              & $1.596$ & (-3.1415,3.1415)  & $1.598_{-0.002}^{+0.003}$ & $1.599_{-0.003}^{+0.003}$ & $1.599_{-0.005}^{+0.005}$ & $1.38_{-0.02}^{+0.02}$ \\ 
\hline                                   %inserts single line
& $\phantom{ }\chi^2_{\rm len,red}$ &  &   & 1.02& 1.02& 1.02 & 1.07 \\
& $\phantom{ }\chi^2_{\rm dyn,red}$ &  &   & 1.1& 0.95 &  & 1.63 \\
& $\phantom{ }\chi^2_{\rm red}$ &  &   & 1.02& 1.02& 1.02 & 1.07 \\
\hline
\end{tabular}
\end{table*}

\begin{table*}
\caption{Modelled lens and source parameter values for the different configurations of the composite mock data (presented in Section \ref{subsubsec:Compositemodels}), with kinematic data having a signal-to-noise of 15 in the brightest pixel and binned to have a signal-to-noise of 30 in each bin (shown in Figure \ref{fig:mock_kin_SN30_comp}). The model presented are that including both the lensing and dynamics constraints (LD composite), that including only the lensing ones (L composite), and the two power-law models, with lensing and dynamics constraints (LD power-law) and with those from lensing only (L power-law). Parameters are presented with the $1\sigma$ uncertainties.}
% title of Table
\label{tab:ModelParaComposite}      % is used to refer this table in the text
\centering                          % used for centering table
\renewcommand{\arraystretch}{1.2}  % change the cell height (row spacing)
\begin{tabular}{llcccccc}        % centred columns (3 columns)
\hline                 % inserts double horizontal lines
\hline
Parameter description & Parameter  &  Input & Prior   & LD composite & L composite & LD power-law & L power-law\\    % table heading 
\hline
\hline                        % inserts single horizontal line
%\hline
 & Source &  &   &   &   &  & \\
\hline
 & light &  &   &   &   &  & \\
x centroid & $\phantom{ }x_{\rm c,s}$     $['']$                 &    $4.1058$ & (2.5,5.5)  & $4.1058_{-0.0003}^{+0.0003}$  & $4.104_{-0.009}^{+0.009}$ & $4.1870_{-0.0002}^{+0.0002}$  & $4.240_{-0.005}^{+0.004}$\\ 
y centroid & $\phantom{ }y_{\rm c,s}$       $['']$               &   $3.4667$  &  (2.,5.) &  $3.4607_{-0.0002}^{+0.0002}$ & $3.468_{-0.008}^{+0.01}$ & $3.4836_{-0.0002}^{+0.0002}$ & $3.459_{-0.004}^{+0.004}$ \\ 
axis ratio & $\phantom{ }q_{\rm s}$                      &$0.850$  & (0.5,1.)  & $0.849_{-0.009}^{+0.01}$ & $0.851_{-0.009}^{+0.01}$  & $0.922_{-0.007}^{+0.008}$ & $0.905_{-0.009}^{+0.009}$\\ 
postion angle & $\phantom{ }\theta_{\rm s}$     $[radians]$                 & $0.00$  & -  & $-0.0028_{-0.03}^{+0.03}$ & $0.00_{-0.03}^{+0.03}$ & $-0.64_{-0.05}^{+0.05}$ & $-0.39_{-0.04}^{+0.04}$ \\ 
intensity & $\phantom{ }I_{\rm e,s}$        $[counts]$              &$200.$  & (120,250)  & $196._{-8.}^{+10.}$ & $195._{-10.}^{+10.}$  & $138._{-8.}^{+6.}$ & $191._{-9.}^{+10.}$\\ 
effective radius & $\phantom{ }R_{\rm eff,s}$                      &$0.5$ & (0.2,2.)  &  $0.51_{-0.01}^{+0.01}$ & $0.51_{-0.02}^{+0.02}$  & $0.62_{-0.02}^{+0.02}$ & $0.52_{-0.02}^{+0.02}$\\ 
Sersic index & $\phantom{ }n_{\rm s}$                      &$4.00$ & (0.5,10.)  & $4.05_{-0.07}^{+0.06}$ & $4.06_{-0.07}^{+0.08}$ & $4.62_{-0.06}^{+0.08}$ & $4.12_{-0.08}^{+0.07}$ \\ 
\hline
 & mass &  &   &   &  &  \\
anisotropy & $\phantom{ }\beta$                      & $-0.15$    & (-0.3,0.3)  & $-0.16_{-0.06}^{+0.05}$ & $-0.17_{-0.06}^{+0.07}$& $-0.15_{-0.07}^{+0.06}$ & $-0.00_{-0.2}^{+0.2}$ \\ 
inclination & $\phantom{ }i$                      & $0.10$  & (0.,0.2))  & $0.09_{-0.06}^{+0.07}$ & $0.04_{-0.03}^{+0.06}$ & $0.10_{-0.07}^{+0.07}$ & $0.10_{-0.07}^{+0.07}$\\ 
axis ratio & $\phantom{ }q^{\rm piemd}_{\rm s}$                      & $0.90$   & (0.5,1.)  & $0.90_{-0.05}^{+0.05}$ & $0.88_{-0.07}^{+0.07}$ & $0.75_{-0.17}^{+0.11}$ & $0.8_{-0.2}^{+0.2}$\\
Einstein radius & $\phantom{ }\theta^{\rm piemd}_{\rm E,\infty,s}$    $['']$                 & $1.15$   & (0.,5)  & $1.15_{-0.03}^{+0.04}$ & $1.56_{-0.08}^{+0.09}$ & $1.29_{-0.07}^{+0.08}$ & $1.6_{-1.1}^{+1.6}$\\
\hline
 & Lens &  &   &   &   &  &  \\
\hline
 & light &  &   &   &   &  &  \\
axis ratio & $\phantom{ }q_{\rm l}$                      & $0.768$ & (0.2,1.)  & $0.769_{-0.003}^{+0.002}$ & $0.770_{-0.003}^{+0.003}$ & $0.772_{-0.003}^{+0.002}$ & $0.773_{-0.003}^{+0.003}$\\ 
intensity & $\phantom{ }I_{\rm e,l}$                       & $965.$  & (800.,1100.)  & $957._{-22.}^{+26.}$ & $958._{-24.}^{+25.}$ & $985._{-27.}^{+27.}$ & $930._{-22.}^{+22.}$\\ 
effective radius & $\phantom{ }R_{\rm eff,l}$                      & $0.90$ & (0.3,2.)  & $0.90_{-0.01}^{+0.01}$ & $0.90_{-0.01}^{+0.01}$ & $0.881_{-0.013}^{+0.014}$ & $0.91_{-0.01}^{+0.01}$\\ 
Sersic index & $\phantom{ }n_{\rm l}$                      &$4.00$  & (0.5,10.)  & $4.03_{-0.05}^{+0.04}$ & $4.03_{-0.04}^{+0.05}$ & $4.03_{-0.05}^{+0.05}$ & $4.09_{-0.04}^{+0.04}$ \\ 
\hline
 & mass (composite) &  &   &   &   \\
x centroid & $\phantom{ }x_{\rm c,l}$       $['']$               &  $4.4108$    &(4.,5.)   & $4.4109_{-0.0005}^{+0.0004}$ & $4.4108_{-0.0004}^{+0.0004}$ & - & - \\ 
y centroid & $\phantom{ }y_{\rm c,l}$        $['']$            &  $4.0112$ &  (3.5,5.5) & $4.0114_{-0.0003}^{+0.0004}$ & $4.0115_{-0.0004}^{+0.0004}$ & - & - \\ 
axis ratio & $\phantom{ }q_{\rm l,1}$                      &$0.883$ & (0.2,1.)  &  $0.884_{-0.007}^{+0.008}$ & $0.88_{-0.01}^{+0.01}$ & - & - \\ 
position angle & $\phantom{ }\theta_{\rm l}$      $[radians]$                &$0.551$  & -  & $0.559_{-0.006}^{+0.006}$ & $0.561_{-0.007}^{+0.007}$ & - & - \\
chameleon parameter &$w_{\rm c,1}$ $['']$ & $2.03$ & (0,10) & $2.03_{-0.04}^{+0.05}$& $2.04_{-0.09}^{+0.08}$ &- & -  \\
chameleon parameter &$w_{\rm t,1}$ $['']$ & $2.47$ & (0,10) & $2.48_{-0.04}^{+0.03}$& $2.48_{-0.07}^{+0.06}$ &- & - \\
axis ratio & $\phantom{ }q_{\rm l,2}$                      &$0.847$ & (0.2,1.)  &  $0.846_{-0.005}^{+0.006}$ & $0.849_{-0.008}^{+0.009}$ & - & - \\ 
chameleon parameter &$w_{\rm c,2}$ $['']$ & $0.06$ & (0,10) & $0.07_{-0.01}^{+0.01}$& $0.06_{-0.02}^{+0.02}$ &- & -  \\
chameleon parameter &$w_{\rm t,2}$ $['']$ & $0.67$ & (0,10) & $0.67_{-0.03}^{+0.03}$& $0.67_{-0.05}^{+0.05}$ &- & -  \\
x centroid & $\phantom{ }x^{\rm nfw}_{\rm E,l}$    $['']$                 & $4.39$ & (3.5,5.)  & $4.36_{-0.04}^{+0.04}$ & $4.35_{-0.04}^{+0.04}$ & - & - \\ 
y centroid & $\phantom{ }y^{\rm nfw}_{\rm l}$                      & $3.90$ & (3.5,5.)  &  $3.89_{-0.02}^{+0.02}$ & $3.88_{-0.02}^{+0.02}$ & - & - \\ 
axis ratio & $\phantom{ }q^{\rm nfw}_{\rm l}$                      &$0.734$ & (0.3,1.)  & $0.733_{-0.006}^{+0.007}$ & $0.737_{-0.009}^{+0.010}$ & - & - \\ 
position angle & $\phantom{ }\theta^{\rm nfw}_{\rm l}$                      &$3.593$ & - & $3.589_{-0.006}^{+0.007}$ & $3.58_{-0.01}^{+0.01}$ & - & - \\ 
Einstein radius & $\phantom{ }\theta^{\rm nfw}_{\rm E,\infty,l}$                      &$0.399$  & (0.,5.) & $0.40_{-0.01}^{+0.01}$ & $0.41_{-0.02}^{+0.02}$ & - & - \\ 
scale radius & $\phantom{ }r^{\rm nfw}_{\rm s,l}$                      &$23.7$  & Gaussian(18.6,2.6)  & $22.8_{-0.9}^{+1.4}$ & $21.3_{-1.2}^{+1.4}$ & - & - \\ 
mass-to-light ratio & $\phantom{ }\frac{M}{L}$             & $4.10$ & (2.5,6.5)  & $4.11_{-0.05}^{+0.05}$ & $4.12_{-0.05}^{+0.05}$ & - & - \\ 
\hline
 & mass (power law) &  &   &   &   &   \\
x centroid & $\phantom{ }x^{\rm spemd}_{\rm c,l}$       $['']$                  &  -   & (4.,5.) & - & - & $4.4105_{-0.0004}^{+0.0005}$ & $4.4109_{-0.0004}^{+0.0004}$ \\ 
y centroid & $\phantom{ }y^{\rm spemd}_{\rm c,l}$        $['']$            &  - & (3.5,5.5)  & - & - & $4.0119_{-0.0004}^{+0.0005}$ & $4.0109_{-0.0004}^{+0.0004}$\\ 
axis ratio & $\phantom{ }q^{\rm spemd}_{\rm l}$                      & - & (0.2,1.)  & - & - &  $0.7179_{-0.001}^{+0.002}$ & $0.736_{-0.002}^{+0.003}$ \\ 
position angle & $\phantom{ }\theta^{\rm spemd}_{\rm l}$      $[radians]$                & - & -  & - & - &  $3.635_{-0.002}^{+0.002}$ & $3.667_{-0.003}^{+0.003}$ \\ 
Einstein radius & $\phantom{ }\theta^{\rm spemd}_{\rm E,l}$    $['']$                 & - & -  & - & - & $3.618_{-0.009}^{+0.008}$ & $3.60_{-0.01}^{+0.02}$ \\ 
slope & $\phantom{ }\gamma^{\rm spemd}_{\rm l}$                      & - & (0.2,0.8)  & - & - & $0.510_{-0.002}^{+0.002}$ & $0.515_{-0.005}^{+0.004}$\\ 
\hline  
 & external shear &  &   &   &   \\
strength & $\phantom{ }\gamma_{\rm ext, \infty}$    $['']$                  &$0.151$ & (0.,0.3) &  $0.152_{-0.001}^{+0.001}$ & $0.150_{-0.004}^{+0.004}$ & $0.171_{-0.0004}^{+0.0004}$ & $0.189_{-0.003}^{+0.005}$ \\ 
position angle & $\phantom{ }\phi_{\rm ext}$        $[radians]$              & $1.416$ & (-3.1415,3.1415) & $1.425_{-0.01}^{+0.01}$ & $1.43_{-0.01}^{+0.01}$& $1.491_{-0.003}^{+0.003}$ & $1.5118_{-0.005}^{+0.005}$ \\
\hline  
\hline                                   %inserts single line
 & $\phantom{ }\chi^2_{\rm len,red}$ &  &   & 0.99& 0.99 & 1.03 & 1.02  \\
  & $\phantom{ }\chi^2_{\rm dyn,red}$ &  &   & 0.99&  & 1.25 &  \\
 & $\phantom{ }\chi^2_{\rm red}$ &  &   & 0.99& 0.99 & 1.03 & 1.02  \\
\hline
\end{tabular}
\end{table*}

\section{Summary}
\label{sec:conclude}

GLaD is a software that performs a joint gravitational lensing and dynamical modelling analysis, with the goal of studying properties of source galaxies and to assess the improvement on the constraints on the lens mass model when high-spatially resolved kinematic observations of the lensed source are avail- able. GLaD assumes parametrised mass and light profiles for both the source and the lens and produces a kinematic map (employing Jeans Anisotropic MGE) and a surface brightness map on the source plane. Then, by computing the deflection produced by the lens mass profile, these quantities are mapped to the image plane. We therefore obtain a lensed kinematic map (binned, if binning is performed on the input data) and a surface brightness map that we then compare to the input kinematic and surface brightness data. We show that this method allows one to infer the mass properties of the source and also to improve constraints on the lens mass parameters. \\
We test this method by simulating two different sets of mock data of the \ourlens\ system, and we remodel all the parameters with different assumptions. The first set of mock data assumes a power-law profile for the lens mass plus external shear, including both the baryonic and dark matter. The second set assumes a composite mass model for the lens mass, i.e. uses a chameleon light profile scaled by a mass-to-light ratio to describe the baryonic mass and a NFW profile for the dark matter component. With the first mock data set we test the improvement on the constraints on both lens and source due to the addition of the kinematic data as compared to a lensing-only analysis. We compare different quality of kinematic data and we test systematics on the source mass model by modelling it with different profiles. Finally we perform the previous analysis using the second mock data set. In this case, we also test systematic errors on the lens mass model, by remodelling the mass with a single power-law profile. Our findings are summarised below:
\begin{itemize}
\item Our software is consistently able to recover the input parameter of the model within the $1-2 \sigma$ uncertainties when modelling with the input configuration set-up, both in the case of the combined lensing and dynamics analysis and in the lensing-only analysis. This is true for both sets of mock data presented.
\item The addition of the source kinematics data allows us to put constraints on the source mass, that is otherwise unconstrained. Moreover, the joint lensing and dynamics analysis tightens constraints on the source centroid and the source light axis ratio (up to a factor $\sim$$20$). Furthermore, the statistical uncertainties on the lens mass axis ratio, Einstein radius and slope are improved by up to a factor $\sim$$3$. Tighter constraints are also observed on the shear parameters (up to a factor $\sim$$5$). 
\item The addition of dynamics allows us to identify degeneracies between lens mass parameters and source kinematic properties. In particular, we note a tight correlation between the source centroid and the lens slope and Einstein radius.
\item Different kinematic data binning do not show, for this particular case, prominent differences on the lens mass parameters, probably due to the subdominant amount of data points with respect to the lensing constraints. Despite this, we do note a factor $\sim1.1-1.2$ improvement on some source parameters, such as \sersic\ amplitude $I_{\rm e,s}$, effective radius $R_{\rm eff,s}$, \sersic\ index $n_{\rm s}$, source's mass axis ratio $q^{\rm piemd}_{\rm s}$ and Einstein radius $\theta^{\rm piemd}_{\rm E,s}$. The slight improvement of SN20 shows that, to tighten parameters' constraints, it is better to obtain more spatially resolved elements, at the expense of higher S/N per bin.
\item When testing systematics, i.e. modelling the source mass with a mass-follows-light profile, we find that this biases the model parameters, in particular the predicted source inclination and the lens's Einstein radius, slope and axis ratio. It also mildly underestimates the source $\varv_{\rm rms}$, leading to a higher $\chi^2_{\rm dyn}$ for this model. The bias is particularly noticeable on the reconstructed surface mass density of the source galaxy which, for this particular system, is underestimated with respect to the data in the regions larger than $0.2\ R_{\rm eff}$. Since the region within $0.2\ R_{\rm eff}$ corresponds to the higher signal-to-noise regions on the image plane, this shows that the surface mass density of the source is very sensitive to the choice of the model in regions that have lower signal-to-noise data. We conclude that the addition of the kinematic analysis can help better distinguish between models than the lensing analysis alone.
\item The analysis carried out with the mock data from the composite mass distribution for the lens mass shows similar results as the previous set of mock data. We still find improved constraints on the source centroid, and also on the dark matter (NFW profile) orientation and Einstein radius, when including the kinematic data compared to the lensing-only analysis. Furthermore, comparing the lensing and dynamics and lensing-only power-law model we find, consistent with the previous set of mock data, that the lens Einstein radius is better constrained by a factor 2 when including dynamics. Another interesting observation from this test is that if we model the composite lens mass with a  single power-law profile, the combined lensing and dynamics analysis has a higher $\chi^2$ than the lensing only. This increased value comes from the misfit of the kinematic data, showing that the addition of these constraints can actually discern better between models.
\end{itemize}
Our analysis shows that the combination of the lensing and the dynamical analysis as complementary probes grants an interesting gain of information on both the source and lens galaxies, and it also allows us to break some of the degeneracies relative to each method. With the improvement in data quality brought by instruments as those on $JWST$ and the $ELT$ ($Extremely\ Large\ Telescope$), this method allows a detailed analysis of the source galaxies, giving us an extensive look into the high-redshift Universe. 

\begin{acknowledgements}
 We thank Michele Cappellari for making JamPy and MGEfit routines publicly available.  GC thanks Stefan Schuldt and Elisa Ritondale for their help and the useful discussions.  GC, AY and SHS thank the Max Planck Society for support through the Max Planck Research Group for SHS.
\end{acknowledgements}

\bibliographystyle{aa}
\bibliography{GLaD}

\begin{thebibliography}{108}
\expandafter\ifx\csname natexlab\endcsname\relax\def\natexlab#1{#1}\fi

\bibitem[{{Auger} {et~al.}(2010){Auger}, {Treu}, {Bolton}, {Gavazzi},
  {Koopmans}, {Marshall}, {Moustakas}, \& {Burles}}]{Auger2010}
{Auger}, M.~W., {Treu}, T., {Bolton}, A.~S., {et~al.} 2010, \apj, 724, 511

\bibitem[{{Bacon} {et~al.}(1995){Bacon}, {Adam}, {Baranne}, {Courtes}, {Dubet},
  {Dubois}, {Emsellem}, {Ferruit}, {Georgelin}, {Monnet}, {Pecontal},
  {Rousset}, \& {Say}}]{Bacon1995}
{Bacon}, R., {Adam}, G., {Baranne}, A., {et~al.} 1995, \aaps, 113, 347

\bibitem[{{Bagnasco} {et~al.}(2007){Bagnasco}, {Kolm}, {Ferruit}, {Honnen},
  {Koehler}, {Lemke}, {Maschmann}, {Melf}, {Noyer}, {Rumler}, {Salvignol},
  {Strada}, \& {Te Plate}}]{Bagnasco2007}
{Bagnasco}, G., {Kolm}, M., {Ferruit}, P., {et~al.} 2007, in Society of
  Photo-Optical Instrumentation Engineers (SPIE) Conference Series, Vol. 6692,
  Cryogenic Optical Systems and Instruments XII, 66920M

\bibitem[{{Baldry} {et~al.}(2004){Baldry}, {Glazebrook}, {Brinkmann},
  {Ivezi{\'c}}, {Lupton}, {Nichol}, \& {Szalay}}]{Baldry2004}
{Baldry}, I.~K., {Glazebrook}, K., {Brinkmann}, J., {et~al.} 2004, \apj, 600,
  681

\bibitem[{{Barkana}(1998)}]{Barkana98}
{Barkana}, R. 1998, \apj, 502, 531

\bibitem[{{Barnab{\`e}} {et~al.}(2011){Barnab{\`e}}, {Czoske}, {Koopmans},
  {Treu}, \& {Bolton}}]{Barnabe2011}
{Barnab{\`e}}, M., {Czoske}, O., {Koopmans}, L.~V.~E., {Treu}, T., \& {Bolton},
  A.~S. 2011, \mnras, 415, 2215

\bibitem[{{Barnab{\`e}} {et~al.}(2012){Barnab{\`e}}, {Dutton}, {Marshall},
  {Auger}, {Brewer}, {Treu}, {Bolton}, {Koo}, \& {Koopmans}}]{Barnabe2012}
{Barnab{\`e}}, M., {Dutton}, A.~A., {Marshall}, P.~J., {et~al.} 2012, \mnras,
  423, 1073

\bibitem[{{Barnab{\`e}} \& {Koopmans}(2007)}]{BarnabeKoopmans2007}
{Barnab{\`e}}, M. \& {Koopmans}, L. V.~E. 2007, \apj, 666, 726

\bibitem[{{Barth} {et~al.}(2001){Barth}, {Sarzi}, {Rix}, {Ho}, {Filippenko}, \&
  {Sargent}}]{Barth2001}
{Barth}, A.~J., {Sarzi}, M., {Rix}, H.-W., {et~al.} 2001, \apj, 555, 685

\bibitem[{{Bell} {et~al.}(2003){Bell}, {McIntosh}, {Katz}, \&
  {Weinberg}}]{Bell2003}
{Bell}, E.~F., {McIntosh}, D.~H., {Katz}, N., \& {Weinberg}, M.~D. 2003, \apjs,
  149, 289

\bibitem[{{Bendinelli}(1991)}]{Bendinell1991}
{Bendinelli}, O. 1991, \apj, 366, 599

\bibitem[{{Bezanson} {et~al.}(2011){Bezanson}, {van Dokkum}, {Franx},
  {Brammer}, {Brinchmann}, {Kriek}, {Labb{\'e}}, {Quadri}, {Rix}, {van de
  Sande}, {Whitaker}, \& {Williams}}]{Bezanson2011}
{Bezanson}, R., {van Dokkum}, P.~G., {Franx}, M., {et~al.} 2011, \apjl, 737,
  L31

\bibitem[{{Binney} \& {Tremaine}(1988)}]{Binney&Tremaine1988}
{Binney}, J. \& {Tremaine}, S. 1988, Sky and Telescope, 76, 45

\bibitem[{{Birkmann} {et~al.}(2016){Birkmann}, {Ferruit}, {Rawle}, {Sirianni},
  {Alves de Oliveira}, {B{\"o}ker}, {Giardino}, {L{\"u}tzgendorf}, {Marston},
  {Stuhlinger}, {te Plate}, {Jensen}, {Rumler}, {Dorner}, {Karl}, {Mosner},
  {Wright}, \& {Rapp}}]{Birkmann2016}
{Birkmann}, S.~M., {Ferruit}, P., {Rawle}, T., {et~al.} 2016, in Society of
  Photo-Optical Instrumentation Engineers (SPIE) Conference Series, Vol. 9904,
  Space Telescopes and Instrumentation 2016: Optical, Infrared, and Millimeter
  Wave, 99040B

\bibitem[{{Birrer} {et~al.}(2016){Birrer}, {Amara}, \&
  {Refregier}}]{Birrer2016}
{Birrer}, S., {Amara}, A., \& {Refregier}, A. 2016, \jcap, 8, 020

\bibitem[{{Birrer} {et~al.}(2017){Birrer}, {Amara}, \&
  {Refregier}}]{Birrer2017}
{Birrer}, S., {Amara}, A., \& {Refregier}, A. 2017, \jcap, 5, 037

\bibitem[{{Cappellari}(2002)}]{Cappellari2002}
{Cappellari}, M. 2002, \mnras, 333, 400

\bibitem[{{Cappellari}(2008)}]{Cappellari2008}
{Cappellari}, M. 2008, \mnras, 390, 71

\bibitem[{{Cappellari}(2016)}]{Cappellari2016}
{Cappellari}, M. 2016, \araa, 54, 597

\bibitem[{{Cappellari} \& {Copin}(2003)}]{Cappellari&Copin2003}
{Cappellari}, M. \& {Copin}, Y. 2003, \mnras, 342, 345

\bibitem[{{Cappellari} {et~al.}(2015){Cappellari}, {Romanowsky}, {Brodie},
  {Forbes}, {Strader}, {Foster}, {Kartha}, {Pastorello}, {Pota}, {Spitler},
  {Usher}, \& {Arnold}}]{Cappellari2015}
{Cappellari}, M., {Romanowsky}, A.~J., {Brodie}, J.~P., {et~al.} 2015, \apjl,
  804, L21

\bibitem[{{Cappellari} {et~al.}(2013){Cappellari}, {Scott}, {Alatalo}, {Blitz},
  {Bois}, {Bournaud}, {Bureau}, {Crocker}, {Davies}, {Davis}, {de Zeeuw},
  {Duc}, {Emsellem}, {Khochfar}, {Krajnovi{\'c}}, {Kuntschner}, {McDermid},
  {Morganti}, {Naab}, {Oosterloo}, {Sarzi}, {Serra}, {Weijmans}, \&
  {Young}}]{Cappellari2013}
{Cappellari}, M., {Scott}, N., {Alatalo}, K., {et~al.} 2013, \mnras, 432, 1709

\bibitem[{{Cappellari} {et~al.}(2002){Cappellari}, {Verolme}, {van der Marel},
  {Verdoes Kleijn}, {Illingworth}, {Franx}, {Carollo}, \& {de
  Zeeuw}}]{Cappellari&Verolme2002}
{Cappellari}, M., {Verolme}, E.~K., {van der Marel}, R.~P., {et~al.} 2002,
  \apj, 578, 787

\bibitem[{{Chang} {et~al.}(2013){Chang}, {van der Wel}, {Rix}, {Holden},
  {Bell}, {McGrath}, {Wuyts}, {H{\"a}ussler}, {Barden}, {Faber}, {Mozena},
  {Ferguson}, {Guo}, {Galametz}, {Grogin}, {Kocevski}, {Koekemoer}, {Dekel},
  {Huang}, {Hathi}, \& {Donley}}]{Chang2013}
{Chang}, Y.-Y., {van der Wel}, A., {Rix}, H.-W., {et~al.} 2013, \apj, 773, 149

\bibitem[{{Chen} {et~al.}(2019){Chen}, {Fassnacht}, {Suyu}, {Rusu}, {Chan},
  {Wong}, {Auger}, {Hilbert}, {Bonvin}, {Birrer}, {Millon}, {Koopmans},
  {Lagattuta}, {McKean}, {Vegetti}, {Courbin}, {Ding}, {Halkola}, {Jee},
  {Shajib}, {Sluse}, {Sonnenfeld}, \& {Treu}}]{Chen2019}
{Chen}, G. C.~F., {Fassnacht}, C.~D., {Suyu}, S.~H., {et~al.} 2019, arXiv
  e-prints, arXiv:1907.02533

\bibitem[{{Cimatti} {et~al.}(2008){Cimatti}, {Cassata}, {Pozzetti}, {Kurk},
  {Mignoli}, {Renzini}, {Daddi}, {Bolzonella}, {Brusa}, {Rodighiero},
  {Dickinson}, {Franceschini}, {Zamorani}, {Berta}, {Rosati}, \&
  {Halliday}}]{Cimatti2008}
{Cimatti}, A., {Cassata}, P., {Pozzetti}, L., {et~al.} 2008, \aap, 482, 21

\bibitem[{{Daddi} {et~al.}(2005){Daddi}, {Renzini}, {Pirzkal}, {Cimatti},
  {Malhotra}, {Stiavelli}, {Xu}, {Pasquali}, {Rhoads}, {Brusa}, {di Serego
  Alighieri}, {Ferguson}, {Koekemoer}, {Moustakas}, {Panagia}, \&
  {Windhorst}}]{Daddi2005}
{Daddi}, E., {Renzini}, A., {Pirzkal}, N., {et~al.} 2005, \apj, 626, 680

\bibitem[{{Dai} \& {Guerras}(2018)}]{Dai&Guerras2018}
{Dai}, X. \& {Guerras}, E. 2018, \apjl, 853, L27

\bibitem[{{Dai} {et~al.}(2010){Dai}, {Kochanek}, {Chartas}, {Koz{\l}owski},
  {Morgan}, {Garmire}, \& {Agol}}]{Dai2010}
{Dai}, X., {Kochanek}, C.~S., {Chartas}, G., {et~al.} 2010, \apj, 709, 278

\bibitem[{{de Zeeuw} {et~al.}(2002){de Zeeuw}, {Bureau}, {Emsellem}, {Bacon},
  {Carollo}, {Copin}, {Davies}, {Kuntschner}, {Miller}, {Monnet}, {Peletier},
  \& {Verolme}}]{deZeeuw2002}
{de Zeeuw}, P.~T., {Bureau}, M., {Emsellem}, E., {et~al.} 2002, \mnras, 329,
  513

\bibitem[{{Di Teodoro} {et~al.}(2018){Di Teodoro}, {Grillo}, {Fraternali},
  {Gobat}, {Karman}, {Mercurio}, {Rosati}, {Balestra}, {Caminha}, \&
  {Caputi}}]{DiTeodoro2018}
{Di Teodoro}, E.~M., {Grillo}, C., {Fraternali}, F., {et~al.} 2018, \mnras,
  476, 804

\bibitem[{{Ding} {et~al.}(2017){Ding}, {Treu}, {Suyu}, {Wong}, {Morishita},
  {Park}, {Sluse}, {Auger}, {Agnello}, {Bennert}, \& {Collett}}]{Ding2017}
{Ding}, X., {Treu}, T., {Suyu}, S.~H., {et~al.} 2017, Monthly Notices of the
  Royal Astronomical Society, 472, 90

\bibitem[{{Dutton} {et~al.}(2011){Dutton}, {Brewer}, {Marshall}, {Auger},
  {Treu}, {Koo}, {Bolton}, {Holden}, \& {Koopmans}}]{Dutton2011}
{Dutton}, A.~A., {Brewer}, B.~J., {Marshall}, P.~J., {et~al.} 2011, \mnras,
  417, 1621

\bibitem[{{Dye} {et~al.}(2015){Dye}, {Furlanetto}, {Swinbank}, {Vlahakis},
  {Nightingale}, {Dunne}, {Eales}, {Smail}, {Oteo}, \& {Hunter}}]{Dye2015}
{Dye}, S., {Furlanetto}, C., {Swinbank}, A.~M., {et~al.} 2015, \mnras, 452,
  2258

\bibitem[{{Emsellem} {et~al.}(2007){Emsellem}, {Cappellari}, {Krajnovi{\'c}},
  {van de Ven}, {Bacon}, {Bureau}, {Davies}, {de Zeeuw}, {Falc{\'o}n-Barroso},
  {Kuntschner}, {McDermid}, {Peletier}, \& {Sarzi}}]{Emsellem2007}
{Emsellem}, E., {Cappellari}, M., {Krajnovi{\'c}}, D., {et~al.} 2007, \mnras,
  379, 401

\bibitem[{{Emsellem} {et~al.}(2004){Emsellem}, {Cappellari}, {Peletier},
  {McDermid}, {Bacon}, {Bureau}, {Copin}, {Davies}, {Krajnovi{\'c}},
  {Kuntschner}, {Miller}, \& {de Zeeuw}}]{Emsellem&Cappellari2004}
{Emsellem}, E., {Cappellari}, M., {Peletier}, R.~F., {et~al.} 2004, \mnras,
  352, 721

\bibitem[{{Emsellem} {et~al.}(1999){Emsellem}, {Dejonghe}, \&
  {Bacon}}]{EmsellemDejonghe&Bacon1999}
{Emsellem}, E., {Dejonghe}, H., \& {Bacon}, R. 1999, \mnras, 303, 495

\bibitem[{{Emsellem} {et~al.}(1994){Emsellem}, {Monnet}, \&
  {Bacon}}]{EmsellemMonnet&Bacon1994A}
{Emsellem}, E., {Monnet}, G., \& {Bacon}, R. 1994, \aap, 285, 723

\bibitem[{{Falc{\'o}n-Barroso} {et~al.}(2017){Falc{\'o}n-Barroso}, {Lyubenova},
  {van de Ven}, {Mendez-Abreu}, {Aguerri}, {Garc{\'{\i}}a-Lorenzo},
  {Bekerait{\'e}}, {S{\'a}nchez}, {Husemann}, {Garc{\'{\i}}a-Benito}, {Mast},
  {Walcher}, {Zibetti}, {Barrera-Ballesteros}, {Galbany},
  {S{\'a}nchez-Bl{\'a}zquez}, {Singh}, {van den Bosch}, {Wild}, {Zhu},
  {Bland-Hawthorn}, {Cid Fernandes}, {de Lorenzo-C{\'a}ceres}, {Gallazzi},
  {Gonz{\'a}lez Delgado}, {Marino}, {M{\'a}rquez}, {P{\'e}rez}, {P{\'e}rez},
  {Roth}, {Rosales-Ortega}, {Ruiz-Lara}, {Wisotzki}, {Ziegler}, \& {Califa
  Collaboration}}]{Barroso2017}
{Falc{\'o}n-Barroso}, J., {Lyubenova}, M., {van de Ven}, G., {et~al.} 2017,
  \aap, 597, A48

\bibitem[{{Foreman-Mackey} {et~al.}(2013){Foreman-Mackey}, {Hogg}, {Lang}, \&
  {Goodman}}]{ForemanMackey16}
{Foreman-Mackey}, D., {Hogg}, D.~W., {Lang}, D., \& {Goodman}, J. 2013, \pasp,
  125, 306

\bibitem[{{Fukugita} {et~al.}(1998){Fukugita}, {Hogan}, \&
  {Peebles}}]{Fukugita1998}
{Fukugita}, M., {Hogan}, C.~J., \& {Peebles}, P.~J.~E. 1998, \apj, 503, 518

\bibitem[{{Gavazzi} {et~al.}(2007){Gavazzi}, {Treu}, {Rhodes}, {Koopmans},
  {Bolton}, {Burles}, {Massey}, \& {Moustakas}}]{Gavazzi2007}
{Gavazzi}, R., {Treu}, T., {Rhodes}, J.~D., {et~al.} 2007, \apj, 667, 176

\bibitem[{{Gebhardt} {et~al.}(2000){Gebhardt}, {Bender}, {Bower}, {Dressler},
  {Faber}, {Filippenko}, {Green}, {Grillmair}, {Ho}, {Kormendy}, {Lauer},
  {Magorrian}, {Pinkney}, {Richstone}, \& {Tremaine}}]{Gebhardt2000}
{Gebhardt}, K., {Bender}, R., {Bower}, G., {et~al.} 2000, \apjl, 539, L13

\bibitem[{{Girard} {et~al.}(2018){Girard}, {Dessauges-Zavadsky}, {Schaerer},
  {Cirasuolo}, {Turner}, {Cava}, {Rodr{\'\i}guez-Mu{\~n}oz}, {Richard}, \&
  {P{\'e}rez-Gonz{\'a}lez}}]{Girard2018}
{Girard}, M., {Dessauges-Zavadsky}, M., {Schaerer}, D., {et~al.} 2018, \aap,
  613, A72

\bibitem[{{Goodman} \& {Weare}(2010)}]{Goodma&Weare2010}
{Goodman}, J. \& {Weare}, J. 2010, Communications in Applied Mathematics and
  Computational Science, 5, 65

\bibitem[{{G{\"u}ltekin} {et~al.}(2009{\natexlab{a}}){G{\"u}ltekin}, {Cackett},
  {Miller}, {Di Matteo}, {Markoff}, \& {Richstone}}]{Gultekin2009b}
{G{\"u}ltekin}, K., {Cackett}, E.~M., {Miller}, J.~M., {et~al.}
  2009{\natexlab{a}}, \apj, 706, 404

\bibitem[{{G{\"u}ltekin} {et~al.}(2012){G{\"u}ltekin}, {Cackett}, {Miller}, {Di
  Matteo}, {Markoff}, \& {Richstone}}]{Gultekin2012}
{G{\"u}ltekin}, K., {Cackett}, E.~M., {Miller}, J.~M., {et~al.} 2012, \apj,
  749, 129

\bibitem[{{G{\"u}ltekin} {et~al.}(2009{\natexlab{b}}){G{\"u}ltekin},
  {Richstone}, {Gebhardt}, {Lauer}, {Pinkney}, {Aller}, {Bender}, {Dressler},
  {Faber}, {Filippenko}, {Green}, {Ho}, {Kormendy}, \&
  {Siopis}}]{Gultekin2009a}
{G{\"u}ltekin}, K., {Richstone}, D.~O., {Gebhardt}, K., {et~al.}
  2009{\natexlab{b}}, \apj, 695, 1577

\bibitem[{{Hogg} {et~al.}(2002){Hogg}, {Blanton}, {Strateva}, {Bahcall},
  {Brinkmann}, {Csabai}, {Doi}, {Fukugita}, {Hennessy}, {Ivezi{\'c}}, {Knapp},
  {Lamb}, {Lupton}, {Munn}, {Nichol}, {Schlegel}, {Schneider}, \&
  {York}}]{Hogg2002}
{Hogg}, D.~W., {Blanton}, M., {Strateva}, I., {et~al.} 2002, \aj, 124, 646

\bibitem[{{Humphrey} \& {Buote}(2010)}]{Humphrey&Buote2010}
{Humphrey}, P.~J. \& {Buote}, D.~A. 2010, \mnras, 403, 2143

\bibitem[{{Jeans}(1922)}]{Jeans1922}
{Jeans}, J.~H. 1922, \mnras, 82, 122

\bibitem[{{Jones} {et~al.}(2010){Jones}, {Swinbank}, {Ellis}, {Richard}, \&
  {Stark}}]{Jones2010}
{Jones}, T.~A., {Swinbank}, A.~M., {Ellis}, R.~S., {Richard}, J., \& {Stark},
  D.~P. 2010, \mnras, 404, 1247

\bibitem[{{Kassiola} \& {Kovner}(1993)}]{KassiolaKovner93}
{Kassiola}, A. \& {Kovner}, I. 1993, \apj, 417, 450

\bibitem[{{Koopmans} {et~al.}(2009){Koopmans}, {Bolton}, {Treu}, {Czoske},
  {Auger}, {Barnab{\`e}}, {Vegetti}, {Gavazzi}, {Moustakas}, \&
  {Burles}}]{Koopmans2009}
{Koopmans}, L.~V.~E., {Bolton}, A., {Treu}, T., {et~al.} 2009, \apjl, 703, L51

\bibitem[{{Koopmans} {et~al.}(2006){Koopmans}, {Treu}, {Bolton}, {Burles}, \&
  {Moustakas}}]{KoopmansEtal06}
{Koopmans}, L.~V.~E., {Treu}, T., {Bolton}, A.~S., {Burles}, S., \&
  {Moustakas}, L.~A. 2006, \apj, 649, 599

\bibitem[{{Kriek} {et~al.}(2008){Kriek}, {van der Wel}, {van Dokkum}, {Franx},
  \& {Illingworth}}]{Kriek2008}
{Kriek}, M., {van der Wel}, A., {van Dokkum}, P.~G., {Franx}, M., \&
  {Illingworth}, G.~D. 2008, \apj, 682, 896

\bibitem[{{Kriek} {et~al.}(2006){Kriek}, {van Dokkum}, {Franx}, {Quadri},
  {Gawiser}, {Herrera}, {Illingworth}, {Labb{\'e}}, {Lira}, {Marchesini},
  {Rix}, {Rudnick}, {Taylor}, {Toft}, {Urry}, \& {Wuyts}}]{Kriek2006}
{Kriek}, M., {van Dokkum}, P.~G., {Franx}, M., {et~al.} 2006, \apjl, 649, L71

\bibitem[{{Kriek} {et~al.}(2009){Kriek}, {van Dokkum}, {Labb{\'e}}, {Franx},
  {Illingworth}, {Marchesini}, \& {Quadri}}]{Kriek2009}
{Kriek}, M., {van Dokkum}, P.~G., {Labb{\'e}}, I., {et~al.} 2009, \apj, 700,
  221

\bibitem[{{Kronawitter} {et~al.}(2000){Kronawitter}, {Saglia}, {Gerhard}, \&
  {Bender}}]{Kronawitter2000}
{Kronawitter}, A., {Saglia}, R.~P., {Gerhard}, O., \& {Bender}, R. 2000, \aaps,
  144, 53

\bibitem[{{Maller} {et~al.}(2000){Maller}, {Simard}, {Guhathakurta}, {Hjorth},
  {Jaunsen}, {Flores}, \& {Primack}}]{Maller2000}
{Maller}, A.~H., {Simard}, L., {Guhathakurta}, P., {et~al.} 2000, \apj, 533,
  194

\bibitem[{{McConnell} {et~al.}(2012){McConnell}, {Ma}, {Murphy}, {Gebhardt},
  {Lauer}, {Graham}, {Wright}, \& {Richstone}}]{McConnell2012}
{McConnell}, N.~J., {Ma}, C.-P., {Murphy}, J.~D., {et~al.} 2012, \apj, 756, 179

\bibitem[{{Monnet} {et~al.}(1992){Monnet}, {Bacon}, \&
  {Emsellem}}]{MonnetBacon&Emsellem1992}
{Monnet}, G., {Bacon}, R., \& {Emsellem}, E. 1992, \aap, 253, 366

\bibitem[{{Morgan} {et~al.}(2006){Morgan}, {Kochanek}, {Falco}, \&
  {Dai}}]{Morgan2006}
{Morgan}, N.~D., {Kochanek}, C.~S., {Falco}, E.~E., \& {Dai}, X. 2006, in
  Bulletin of the American Astronomical Society, Vol.~38, American Astronomical
  Society Meeting Abstracts, 927

\bibitem[{{Murphy} {et~al.}(2011){Murphy}, {Gebhardt}, \& {Adams}}]{Murphy2011}
{Murphy}, J.~D., {Gebhardt}, K., \& {Adams}, J.~J. 2011, \apj, 729, 129

\bibitem[{{Naab} {et~al.}(2014){Naab}, {Oser}, {Emsellem}, {Cappellari},
  {Krajnovi{\'c}}, {McDermid}, {Alatalo}, {Bayet}, {Blitz}, {Bois}, {Bournaud},
  {Bureau}, {Crocker}, {Davies}, {Davis}, {de Zeeuw}, {Duc}, {Hirschmann},
  {Johansson}, {Khochfar}, {Kuntschner}, {Morganti}, {Oosterloo}, {Sarzi},
  {Scott}, {Serra}, {van de Ven}, {Weijmans}, \& {Young}}]{Naab2014}
{Naab}, T., {Oser}, L., {Emsellem}, E., {et~al.} 2014, \mnras, 444, 3357

\bibitem[{{Navarro} {et~al.}(1997){Navarro}, {Frenk}, \& {White}}]{Navarro1997}
{Navarro}, J.~F., {Frenk}, C.~S., \& {White}, S. D.~M. 1997, \apj, 490, 493

\bibitem[{{Newman} {et~al.}(2018){Newman}, {Belli}, {Ellis}, \&
  {Patel}}]{Newman2018}
{Newman}, A.~B., {Belli}, S., {Ellis}, R.~S., \& {Patel}, S.~G. 2018, \apj,
  862, 126

\bibitem[{{Newman} {et~al.}(2017){Newman}, {Smith}, {Conroy}, {Villaume}, \&
  {van Dokkum}}]{Newman2017}
{Newman}, A.~B., {Smith}, R.~J., {Conroy}, C., {Villaume}, A., \& {van Dokkum},
  P. 2017, \apj, 845, 157

\bibitem[{{Oldham} {et~al.}(2017){Oldham}, {Auger}, {Fassnacht}, {Treu},
  {Brewer}, {Koopmans}, {Lagattuta}, {Marshall}, {McKean}, \&
  {Vegetti}}]{Oldham2017}
{Oldham}, L., {Auger}, M.~W., {Fassnacht}, C.~D., {et~al.} 2017, \mnras, 465,
  3185

\bibitem[{{Oser} {et~al.}(2010){Oser}, {Ostriker}, {Naab}, {Johansson}, \&
  {Burkert}}]{2010ApJ...725.2312O}
{Oser}, L., {Ostriker}, J.~P., {Naab}, T., {Johansson}, P.~H., \& {Burkert}, A.
  2010, \apj, 725, 2312

\bibitem[{{Patr{\'\i}cio} {et~al.}(2018){Patr{\'\i}cio}, {Richard}, {Carton},
  {Contini}, {Epinat}, {Brinchmann}, {Schmidt}, {Krajnovi{\'c}}, {Bouch{\'e}},
  \& {Weilbacher}}]{Patricio2018}
{Patr{\'\i}cio}, V., {Richard}, J., {Carton}, D., {et~al.} 2018, \mnras, 477,
  18

\bibitem[{{Rizzo} {et~al.}(2018){Rizzo}, {Vegetti}, {Fraternali}, \& {Di
  Teodoro}}]{Rizzo2018}
{Rizzo}, F., {Vegetti}, S., {Fraternali}, F., \& {Di Teodoro}, E. 2018, \mnras,
  481, 5606

\bibitem[{{Rybak} {et~al.}(2015){Rybak}, {Vegetti}, {McKean}, {Andreani}, \&
  {White}}]{Rybak2015}
{Rybak}, M., {Vegetti}, S., {McKean}, J.~P., {Andreani}, P., \& {White},
  S.~D.~M. 2015, \mnras, 453, L26

\bibitem[{{Sarzi} {et~al.}(2001){Sarzi}, {Rix}, {Shields}, {Rudnick}, {Ho},
  {McIntosh}, {Filippenko}, \& {Sargent}}]{Sarzi2001}
{Sarzi}, M., {Rix}, H.-W., {Shields}, J.~C., {et~al.} 2001, \apj, 550, 65

\bibitem[{{Sluse} {et~al.}(2007){Sluse}, {Claeskens}, {Hutsem{\'e}kers}, \&
  {Surdej}}]{Sluse2007}
{Sluse}, D., {Claeskens}, J.~F., {Hutsem{\'e}kers}, D., \& {Surdej}, J. 2007,
  \aap, 468, 885

\bibitem[{{Sluse} {et~al.}(2003){Sluse}, {Surdej}, {Claeskens},
  {Hutsem{\'e}kers}, {Jean}, {Courbin}, {Nakos}, {Billeres}, \&
  {Khmil}}]{Sluse2003}
{Sluse}, D., {Surdej}, J., {Claeskens}, J.-F., {et~al.} 2003, \aap, 406, L43

\bibitem[{{Suyu} {et~al.}(2013){Suyu}, {Auger}, {Hilbert}, {Marshall}, {Tewes},
  {Treu}, {Fassnacht}, {Koopmans}, {Sluse}, {Blandford}, {Courbin}, \&
  {Meylan}}]{Suyu2013}
{Suyu}, S.~H., {Auger}, M.~W., {Hilbert}, S., {et~al.} 2013, \apj, 766, 70

\bibitem[{{Suyu} \& {Halkola}(2010)}]{SuyuHalkola2010}
{Suyu}, S.~H. \& {Halkola}, A. 2010, \aap, 524, A94

\bibitem[{{Suyu} {et~al.}(2012){Suyu}, {Hensel}, {McKean}, {Fassnacht}, {Treu},
  {Halkola}, {Norbury}, {Jackson}, {Schneider}, {Thompson}, {Auger},
  {Koopmans}, \& {Matthews}}]{Suyu2012a}
{Suyu}, S.~H., {Hensel}, S.~W., {McKean}, J.~P., {et~al.} 2012, \apj, 750, 10

\bibitem[{{Suyu} {et~al.}(2009){Suyu}, {Marshall}, {Blandford}, {Fassnacht},
  {Koopmans}, {McKean}, \& {Treu}}]{SuyuEtal09}
{Suyu}, S.~H., {Marshall}, P.~J., {Blandford}, R.~D., {et~al.} 2009, \apj, 691,
  277

\bibitem[{{Suyu} {et~al.}(2006){Suyu}, {Marshall}, {Hobson}, \&
  {Blandford}}]{Suyu2006}
{Suyu}, S.~H., {Marshall}, P.~J., {Hobson}, M.~P., \& {Blandford}, R.~D. 2006,
  \mnras, 371, 983

\bibitem[{{Suyu} {et~al.}(2014){Suyu}, {Treu}, {Hilbert}, {Sonnenfeld},
  {Auger}, {Blandford}, {Collett}, {Courbin}, {Fassnacht}, {Koopmans},
  {Marshall}, {Meylan}, {Spiniello}, \& {Tewes}}]{Suyu2014}
{Suyu}, S.~H., {Treu}, T., {Hilbert}, S., {et~al.} 2014, \apjl, 788, L35

\bibitem[{{Swinbank} {et~al.}(2015){Swinbank}, {Dye}, {Nightingale},
  {Furlanetto}, {Smail}, {Cooray}, {Dannerbauer}, {Dunne}, {Eales}, \&
  {Gavazzi}}]{Swinbank2015}
{Swinbank}, A.~M., {Dye}, S., {Nightingale}, J.~W., {et~al.} 2015, \apjl, 806,
  L17

\bibitem[{{Szomoru} {et~al.}(2012){Szomoru}, {Franx}, \& {van
  Dokkum}}]{SzomoruFranxvanDokkum2012}
{Szomoru}, D., {Franx}, M., \& {van Dokkum}, P.~G. 2012, \apj, 749, 121

\bibitem[{{Szomoru} {et~al.}(2010){Szomoru}, {Franx}, {van Dokkum}, {Trenti},
  {Illingworth}, {Labb{\'e}}, {Bouwens}, {Oesch}, \& {Carollo}}]{Szomoru2010}
{Szomoru}, D., {Franx}, M., {van Dokkum}, P.~G., {et~al.} 2010, \apjl, 714,
  L244

\bibitem[{{Thomas} {et~al.}(2016){Thomas}, {Ma}, {McConnell}, {Greene},
  {Blakeslee}, \& {Janish}}]{Thomas2016}
{Thomas}, J., {Ma}, C.-P., {McConnell}, N.~J., {et~al.} 2016, \nat, 532, 340

\bibitem[{{Toft} {et~al.}(2012){Toft}, {Gallazzi}, {Zirm}, {Wold}, {Zibetti},
  {Grillo}, \& {Man}}]{Toft2012}
{Toft}, S., {Gallazzi}, A., {Zirm}, A., {et~al.} 2012, \apj, 754, 3

\bibitem[{{Toft} {et~al.}(2007){Toft}, {van Dokkum}, {Franx}, {Labbe},
  {F{\"o}rster Schreiber}, {Wuyts}, {Webb}, {Rudnick}, {Zirm}, {Kriek}, {van
  der Werf}, {Blakeslee}, {Illingworth}, {Rix}, {Papovich}, \&
  {Moorwood}}]{Toft2007}
{Toft}, S., {van Dokkum}, P., {Franx}, M., {et~al.} 2007, \apj, 671, 285

\bibitem[{{Toft} {et~al.}(2005){Toft}, {van Dokkum}, {Franx}, {Thompson},
  {Illingworth}, {Bouwens}, \& {Kriek}}]{Toft2005}
{Toft}, S., {van Dokkum}, P., {Franx}, M., {et~al.} 2005, \apjl, 624, L9

\bibitem[{{Treu} \& {Koopmans}(2002)}]{Treu&Koopmans2002}
{Treu}, T. \& {Koopmans}, L. V.~E. 2002, \apj, 575, 87

\bibitem[{{Treu} \& {Koopmans}(2004)}]{Treu&Koopmans2004}
{Treu}, T. \& {Koopmans}, L. V.~E. 2004, \apj, 611, 739

\bibitem[{{Trujillo} {et~al.}(2006){Trujillo}, {Feulner}, {Goranova}, {Hopp},
  {Longhetti}, {Saracco}, {Bender}, {Braito}, {Della Ceca}, {Drory},
  {Mannucci}, \& {Severgnini}}]{Trujillo2006}
{Trujillo}, I., {Feulner}, G., {Goranova}, Y., {et~al.} 2006, \mnras, 373, L36

\bibitem[{{van de Sande} {et~al.}(2013){van de Sande}, {Kriek}, {Franx}, {van
  Dokkum}, {Bezanson}, {Bouwens}, {Quadri}, {Rix}, \&
  {Skelton}}]{vandeSande2013}
{van de Sande}, J., {Kriek}, M., {Franx}, M., {et~al.} 2013, \apj, 771, 85

\bibitem[{{van de Ven} {et~al.}(2010){van de Ven}, {Falc{\'o}n-Barroso},
  {McDermid}, {Cappellari}, {Miller}, \& {de Zeeuw}}]{vandeVen2010}
{van de Ven}, G., {Falc{\'o}n-Barroso}, J., {McDermid}, R.~M., {et~al.} 2010,
  \apj, 719, 1481

\bibitem[{{van den Bosch} \& {de Zeeuw}(2010)}]{VanDenBosch2010}
{van den Bosch}, R. C.~E. \& {de Zeeuw}, P.~T. 2010, \mnras, 401, 1770

\bibitem[{{van den Bosch} {et~al.}(2012){van den Bosch}, {Gebhardt},
  {G{\"u}ltekin}, {van de Ven}, {van der Wel}, \& {Walsh}}]{VanDenBosch2012}
{van den Bosch}, R.~C.~E., {Gebhardt}, K., {G{\"u}ltekin}, K., {et~al.} 2012,
  \nat, 491, 729

\bibitem[{{van der Wel} {et~al.}(2014){van der Wel}, {Franx}, {van Dokkum},
  {Skelton}, {Momcheva}, {Whitaker}, {Brammer}, {Bell}, {Rix}, {Wuyts},
  {Ferguson}, {Holden}, {Barro}, {Koekemoer}, {Chang}, {McGrath},
  {H{\"a}ussler}, {Dekel}, {Behroozi}, {Fumagalli}, {Leja}, {Lundgren},
  {Maseda}, {Nelson}, {Wake}, {Patel}, {Labb{\'e}}, {Faber}, {Grogin}, \&
  {Kocevski}}]{vanderWel2014}
{van der Wel}, A., {Franx}, M., {van Dokkum}, P.~G., {et~al.} 2014, \apj, 788,
  28

\bibitem[{{van der Wel} {et~al.}(2008){van der Wel}, {Holden}, {Zirm}, {Franx},
  {Rettura}, {Illingworth}, \& {Ford}}]{vanderWel2008}
{van der Wel}, A., {Holden}, B.~P., {Zirm}, A.~W., {et~al.} 2008, \apj, 688, 48

\bibitem[{{van der Wel} {et~al.}(2011){van der Wel}, {Rix}, {Wuyts}, {McGrath},
  {Koekemoer}, {Bell}, {Holden}, {Robaina}, \& {McIntosh}}]{vanderWel2011}
{van der Wel}, A., {Rix}, H.-W., {Wuyts}, S., {et~al.} 2011, \apj, 730, 38

\bibitem[{{van Dokkum} {et~al.}(2008){van Dokkum}, {Franx}, {Kriek}, {Holden},
  {Illingworth}, {Magee}, {Bouwens}, {Marchesini}, {Quadri}, {Rudnick},
  {Taylor}, \& {Toft}}]{vanDokkum2008}
{van Dokkum}, P.~G., {Franx}, M., {Kriek}, M., {et~al.} 2008, \apjl, 677, L5

\bibitem[{{van Dokkum} {et~al.}(2009){van Dokkum}, {Kriek}, \&
  {Franx}}]{vanDokkumKriekFranx2009}
{van Dokkum}, P.~G., {Kriek}, M., \& {Franx}, M. 2009, \nat, 460, 717

\bibitem[{{van Dokkum} {et~al.}(2010){van Dokkum}, {Whitaker}, {Brammer},
  {Franx}, {Kriek}, {Labb{\'e}}, {Marchesini}, {Quadri}, {Bezanson},
  {Illingworth}, {Muzzin}, {Rudnick}, {Tal}, \& {Wake}}]{vanDokkum2010}
{van Dokkum}, P.~G., {Whitaker}, K.~E., {Brammer}, G., {et~al.} 2010, \apj,
  709, 1018

\bibitem[{{Walsh} {et~al.}(2013){Walsh}, {Barth}, {Ho}, \& {Sarzi}}]{Walsh2013}
{Walsh}, J.~L., {Barth}, A.~J., {Ho}, L.~C., \& {Sarzi}, M. 2013, \apj, 770, 86

\bibitem[{{Weijmans} {et~al.}(2009){Weijmans}, {Cappellari}, {Bacon}, {de
  Zeeuw}, {Emsellem}, {Falc{\'o}n-Barroso}, {Kuntschner}, {McDermid}, {van den
  Bosch}, \& {van de Ven}}]{Weijmans2009}
{Weijmans}, A.-M., {Cappellari}, M., {Bacon}, R., {et~al.} 2009, \mnras, 398,
  561

\bibitem[{{Wellons} {et~al.}(2016){Wellons}, {Torrey}, {Ma}, {Rodriguez-Gomez},
  {Pillepich}, {Nelson}, {Genel}, {Vogelsberger}, \&
  {Hernquist}}]{2016MNRAS.456.1030W}
{Wellons}, S., {Torrey}, P., {Ma}, C.-P., {et~al.} 2016, \mnras, 456, 1030

\bibitem[{{Y{\i}ld{\i}r{\i}m} {et~al.}(2019){Y{\i}ld{\i}r{\i}m}, {Suyu}, \&
  {Halkola}}]{Yildirim2019}
{Y{\i}ld{\i}r{\i}m}, A., {Suyu}, S.~H., \& {Halkola}, A. 2019, arXiv e-prints

\bibitem[{{Y{\i}ld{\i}r{\i}m} {et~al.}(2017){Y{\i}ld{\i}r{\i}m}, {van den
  Bosch}, {van de Ven}, {Mart{\'{\i}}n-Navarro}, {Walsh}, {Husemann},
  {G{\"u}ltekin}, \& {Gebhardt}}]{Yildirim2017}
{Y{\i}ld{\i}r{\i}m}, A., {van den Bosch}, R.~C.~E., {van de Ven}, G., {et~al.}
  2017, \mnras, 468, 4216

\bibitem[{{Zirm} {et~al.}(2007){Zirm}, {van der Wel}, {Franx}, {Labb{\'e}},
  {Trujillo}, {van Dokkum}, {Toft}, {Daddi}, {Rudnick}, {Rix},
  {R{\"o}ttgering}, \& {van der Werf}}]{Zirm2007}
{Zirm}, A.~W., {van der Wel}, A., {Franx}, M., {et~al.} 2007, \apj, 656, 66

\end{thebibliography}

\end{document}